\documentclass[aps,prb,twocolumn,groupedaddress,longbibliography]{revtex4-2}

\usepackage[utf8]{inputenc}
\usepackage[T1]{fontenc}
\usepackage{amsmath,amssymb}
\usepackage{physics}
\usepackage{graphicx}
\usepackage{bm}
\usepackage{hyperref}
\usepackage{xcolor}

\newcommand{\de}{\mathrm{d}}

\newcommand{\ie}{{i.e.}}

\begin{document}

\title{Anderson Localization on Husimi Trees and its implications for Many-Body localization}

\author{Dafne Prado Bandeira}
\email{dafne.prado_bandeira@sorbonne-universite.fr}

\author{Marco Tarzia}
\email{marco.tarzia@sorbonne-universite.fr}

\affiliation{Laboratoire de Physique Théorique de la Matière Condensée (LPTMC), CNRS UMR 7600, Sorbonne Université, 4 Place Jussieu, F-75005 Paris, France}

\date{\today}

\begin{abstract}
Motivated by the analogy between many-body localization (MBL) and single-particle Anderson localization on hierarchical graphs, we study localization on the Husimi tree, a generalization of the Bethe lattice with a finite density of local loops of arbitrary but finite length. The exact solution of the model provides a transparent and quantitative framework to systematically inspect the effect of loops on localization. Our analysis indicates that local loops enhance resonant processes, thereby reducing the critical disorder with increasing  their number and size. At the same time, loops promote local hybridization, leading to an increase in the spatial extent of localized eigenstates. These effects reconcile key discrepancies between MBL phenomenology and its single-particle Anderson analog on hierarchical graphs. These results show that local loops are a crucial structural ingredient for realistic single-particle representations of many-body Hilbert spaces.
\end{abstract}

\maketitle

\section{Introduction}
Disorder-induced localization of quantum states plays a fundamental role in understanding quantum transport and ergodicity in disordered systems. In non-interacting systems, Anderson localization (AL) describes the suppression of diffusion by quantum interference in the presence of random potentials~\cite{anderson1958absence,lee1985disordered,evers2008anderson,lagendijk2009fifty}. The extension of this concept to interacting systems defines the phenomenon of many-body localization (MBL), in which disorder and quantum fluctuations combine to prevent thermalization and preserve memory of initial conditions for arbitrary long times even in presence of interactions and for highly excited states~\cite{gornyi2005interacting, basko2006metal,Nandkishore2015,abanin2019MBLcolloquium,alet2018many,sierant2025many,Huse2014}.
Despite extensive analytical and numerical efforts, the stability of the MBL phase in the thermodynamic limit remains under debate: strong finite-size effects complicate any sharp identification of the critical disorder~\cite{suntajs2020quantum,sierant_thouless_2020, kiefer-emmanouilidis_evidence_2020, kiefer-emmanouilidis_slow_2021, sierant_challenges_2022,Beraetal2017, WeinerEversBera2019,EversmodakBera2023,sierant_polynomially_2020,doggen_many-body_2018},
and rare thermal inclusions and rare resonances may destabilize the insulating phase over a very broad disorder range~\cite{deroeck2017stability, szoldra2024catching, sels2023thermalization,  sels_bath-induced_2022,peacock_many-body_2023,leonard_probing_2023,Crowley2020,goihl_exploration_2019,thiery_many-body_2018,luitz_how_2017,morningstar_avalanches_2022,Ha2023,Garratt2021,De_Tomasi2021-ua,colbois_interaction-driven_2024,colbois_statistics_2024,biroli_large-deviation_2024,miranda2025largedeviationsmanybodylocalization,de2024absence} (yet recent rigorous results demonstrate absence of standard diffusion at strong enough disorder in spin chains~\cite{imbrie2016diagonalization, de2024absence}).
The main obstacle to progress lies in the fact that MBL is an inherently non-perturbative, out-of-equilibrium quantum phase transition driven by the interplay of disorder, interactions, and quantum fluctuations, for which no established theoretical framework yet exists~\cite{Nandkishore2015,abanin2019MBLcolloquium,alet2018many,sierant2025many,Huse2014,dalessio2016quantum}.

In this context, a useful perspective on MBL consists in recasting the unitary dynamics as a single-particle hopping in the configuration space, where each node of the resulting graph represents a many-body configuration on the chosen Hilbert-space basis, and the effective hopping encodes the interaction terms connecting them~\cite{basko2006metal,gornyi2005interacting,altshuler1997quasiparticle,tikhonov2021anderson}.
The resulting Hilbert-space graph is effectively high-dimensional, motivating the study of single-particle AL on hierarchical structures as a pictorial and simplified description of MBL in interacting systems~\cite{tikhonov2021anderson,tikhonov2021eigenstate,tarzia2020many,de2013ergodicity,biroli2017delocalized,biroli2020anomalous,logan2019many,garcia2022critical,herre2023ergodicity}. The Bethe lattice is central to this approach: its tree-like structure allows for an exact self-consistent solution of AL, providing a controlled framework to study localization ~\cite{abou1973selfconsistent, evers2008anderson, biroli2010anderson, rizzo2024localized, tikhonov2019statistics, tikhonov2019critical, efetov1985anderson,efetov1987density,efetov1987anderson,zirnbauer1986localization,zirnbauer1986anderson,verbaarschot1988graded,mirlin1991localization,mirlin1991universality,fyodorov1991localization,mirlin1994statistical,mirlin1994distribution,biroli2022critical,roy2024fock}. However, the Bethe lattice only offers a stylized representation of the configuration space of MBL systems.
Hilbert-space graphs generically exhibit a local connectivity that grows linearly with the number of interacting degrees of freedom, contains loops on all scales, as well as correlated disorder, key features absent from the Bethe lattice analogy.

These structural differences may be at the origin of significant discrepancies that arise when MBL is recast as single-particle AL on the Bethe lattice. The first discrepancy concerns the critical disorder strength required to localize single-particle states on the tree ($W_c \! \sim \! z \ln z$, $z$ being the local degree), which is much larger than the effective disorder characterizing the MBL Hilbert space ($W_c \!  \sim \!  \sqrt{z}$). To understand this scaling, consider the paradigmatic models typically used to study MBL: generic disordered, interacting quantum spin chains of $n$ spins~\cite{mace_multifractal_2019,pal2010many,luitz2015many-body,imbrie2016diagonalization,de2024absence,biroli_large-deviation_2024} (see Sec.~\ref{sec:MBL} for a concrete example). In the basis of simultaneous eigenstates of the $\sigma_i^z$ operators, the Hilbert-space graph of such systems is generically a high-dimensional complex graph with local degree $z$ proportional to $n$. Each node corresponds to a spin configuration in this basis, and the on-site energies are sums of $O(n)$ independent random fields $h_i \in [-h, h]$; consequently, the effective diagonal disorder at the MBL transition scales as $W_c^{\rm MBL} \propto \sqrt{n}\, h_c$. The second discrepancy concerns the spatial structure of localized states. 
Localized eigenstates on the Bethe lattice occupy a finite volume, corresponding to a finite inverse participation ratio (IPR)~\cite{rizzo2024localized}, whereas many-body localized eigenstates necessarily extend over a subextensive region of the Hilbert space, resulting in a vanishing IPR in the thermodynamic limit~\cite{mace_multifractal_2019}.
In this context, previous works have addressed these issues by studying how correlations affect the scaling of the critical disorder with system size~\cite{roy2020fock, roy2020localization, scoquart2024role,logan2025multifractality}. 
The impact of local loops, however, has remained unexplored~\cite{saha2026andersonlocalisationspatiallystructured}. To address this gap, we study the  Anderson model on the Husimi tree, a hierarchical graph built from overlapping cliques (fully connected subgraphs) that extends the Bethe lattice by incorporating a controlled density of local loops of arbitrary but finite length. This construction remains analytically tractable, allowing a systematic exploration of how loop topology affects localization. We find that local loops greatly influence the transition, simultaneously lowering the critical disorder and increasing the spatial extent of localized states, resolving both inconsistencies in the single-particle analogy of MBL. At the same time, we show that the universality class of the localization transition remains unchanged from that of the loopless infinite tree~\cite{abou1973selfconsistent,efetov1985anderson,efetov1987density,efetov1987anderson,zirnbauer1986localization,zirnbauer1986anderson,verbaarschot1988graded,mirlin1991localization,mirlin1991universality,fyodorov1991localization,mirlin1994statistical,tikhonov2019statistics,tikhonov2019critical,mirlin1991localization,mirlin1994distribution,biroli2022critical}. 

\section{The model}
\begin{figure}
    \centering
    \includegraphics[width=1\linewidth]{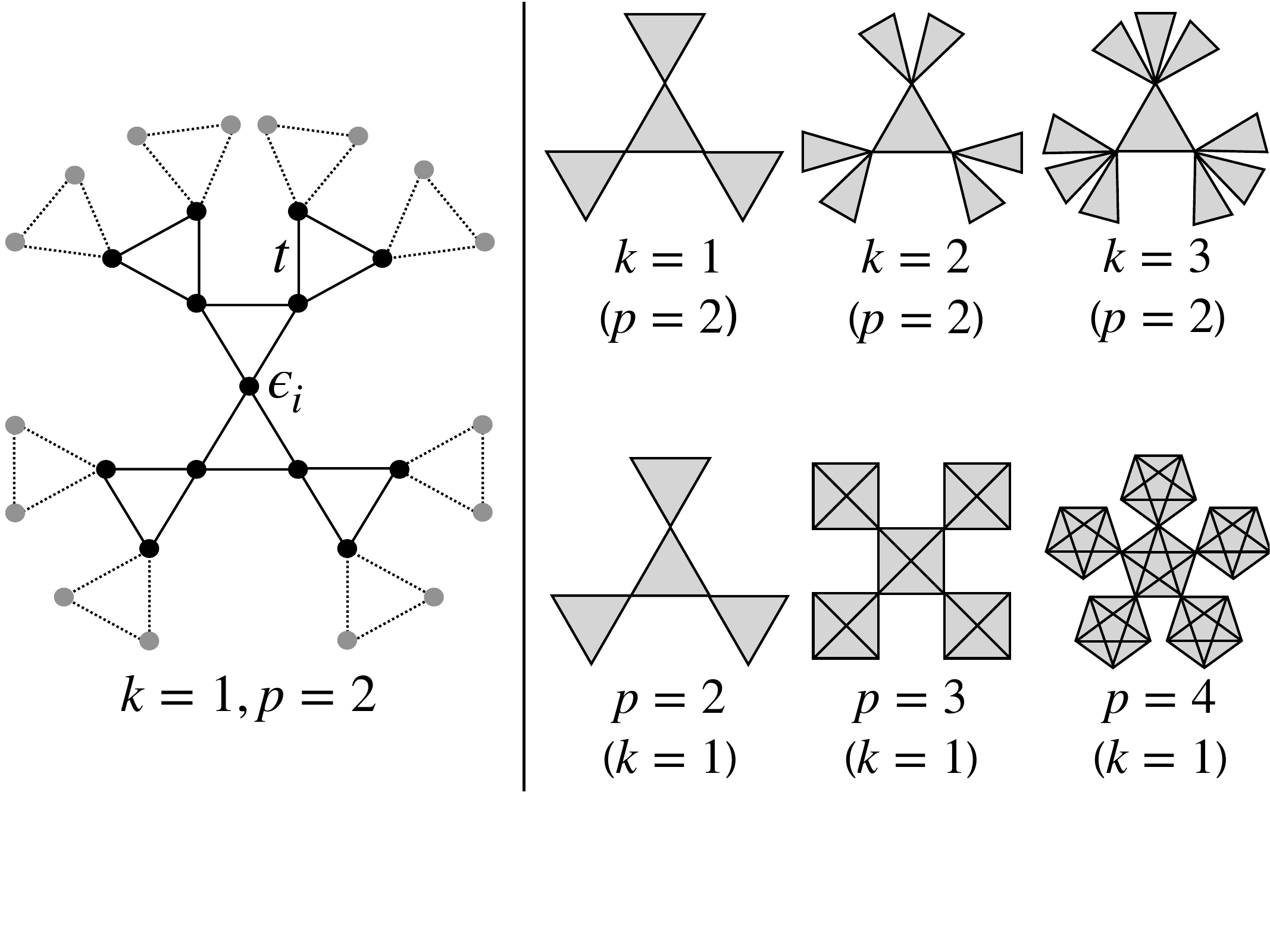}
    \vspace{-1.4cm}
    \caption{Schematic representation of a the Husimi tree: the left panel shows a small portion of the Husimi tree for $k=1$ and $p=2$, while the right panel displays a single clique together with its immediate neighboring cliques for varying values of $k$ at fixed $p=2$ (top) and varying $p$ at fixed $k=1$ (bottom), illustrating the effect of these parameters on the local topology of the graph.}
    \label{fig:husimi_ex} 
\end{figure}

To mimic key structural features of the many-body Hilbert space beyond the Bethe lattice, we study the Anderson model on a Husimi tree, a hierarchical graph in which each node belongs to $k+1$ $(p+1)$--cliques, \ie, fully connected subgraphs of $p+1$ nodes. Neighboring cliques share exactly one node. The local topology is thus specified by the number of cliques per node ($k+1$) and the clique size ($p+1$), giving a node degree $z = (k+1)p$.
A portion of the Husimi tree for $k=1$ and $p=2$ is shown in Fig.~\ref{fig:husimi_ex} (left), while the right panel illustrates how varying $k$ and $p$ shapes the local topology of the graph.

This construction introduces loops of size $3$ to $p+1$, whose number from a given node grows roughly as $(k+1)p^\ell$ [see Eq.~\eqref{eq:loops} and Appendix ~\ref{app:loops} for a derivation of the loop statistics in the Husimi tree], giving an average loop size $\langle\ell\rangle\sim p$. Although this loop distribution differs from that of  Hilbert-space graphs of many-body systems, the Husimi tree provides a controlled framework for studying how increasing the number and the size of local loops affects the properties of AL, while keeping the degree fixed.

The Hamiltonian takes the standard form of a single-particle tight-binding model with on-site disorder and nearest-neighbor hopping,
\begin{align}\label{eq:hamiltonian}
    \hat{H} &= t \sum_{\langle i,j \rangle} \left ( |j\rangle \langle i| + |i\rangle \langle j| \right )
    + \sum_i \epsilon_i |i\rangle \langle i| \nonumber\\
     &= t \sum_{\mu} \! \Bigg( \! \sum_{i\in \partial_\mu} |i\rangle \! \Bigg) \! \Bigg( \! \sum_{i\in \partial_\mu} \langle i| \! \Bigg) \!
    + \sum_i \big [\epsilon_i - (k+1) t \big] |i\rangle \langle i| \, ,
\end{align}
where $t$ is the hopping amplitude between neighboring sites and $\epsilon_i$ are independent random variables, drawn, as customary, from a uniform distribution over $[-W/2,W/2]$. Hereafter, latin indices $i=1, \ldots, N$ label the nodes, and Greek indices $\mu = 1, \ldots, M = (k+1)N/(p+1)$ label the cliques and $i\in \partial_\mu$ denotes all nodes $i$ that belong to clique $\mu$. The second line is a rewriting of $\hat{H}$ in terms of the operators $\sum_{i\in \partial_\mu}|i\rangle$, which define collective clique modes, highlighting that, in the absence of disorder, the Hamiltonian~\eqref{eq:hamiltonian} reduces to a sparse analog  of a (traceless) Wishart matrix~\cite{Wishart1928,potters2020first,kutlin2021emergent} (see Appendix ~\ref{app:zeroW} for more details). Since localization begins at the edge of the band~\cite{abou1974self,biroli2010anderson} (see Fig.~\ref{fig:mobility_edge}), we restrict our analysis to $E=0$ for simplicity. 

The relevant order parameter for AL is the probability distribution of the local density of states (LDoS), $ \rho_i = \sum_\alpha |\psi_\alpha(i)|^2\, \delta(E_\alpha)$.  The LDoS is proportional to the inverse lifetime of a particle created at site $i$, and its typical value reflects the diffusion coefficient~\cite{mirlin1991localization}.  
In the metallic phase, extended states yield a finite typical $\rho_i$, while in the localized regime, $\rho_i$ vanishes in the thermodynamic limit on most of the nodes. The LDoS can be expressed in terms of the Green's functions, $G_{ii} = \langle i | ({\rm i} \eta - \hat{H})^{-1} | i \rangle$, as $\rho_i = \frac{1}{\pi}\lim_{\eta\to0^+}\mathrm{Im}\,G_{ii}$ so that the onset of AL corresponds to the transition of the LDoS distribution from a nonvanishing stable distribution function in the metallic phase to singular behavior as $\eta\!\to\!0^+$ in the insulating phase~\cite{mirlin1991localization,tikhonov2019critical}.
Other spectral observables follow from the same Green's functions, notably the IPR $ I_2 = \big\langle \sum_i |\psi_\alpha(i)|^4 \delta(E_\alpha)\big\rangle = \lim_{\eta \to 0^+} \langle \eta |G_{ii}|^2 \rangle/\langle \mathrm{Im}\,G_{ii} \rangle$, which quantifies the spatial extent of eigenfunctions. $I_2$  converges to a value of order one for localized states,  and vanishes as $1/N$ in the metallic phase~\cite{rizzo2024localized,mirlin1991localization,tikhonov2019statistics}.

\section{Exact solution}
Thanks to the hierarchical structure of the Husimi tree, one can derive exact self-consistent equations for the diagonal elements of the cavity Green’s functions, $G_i^{(\nu)}$, defined as the Green’s function of node $i$ in the absence of the clique $\nu$:
\begin{equation}\label{eq:cavity}
    \left[ G_i^{(\nu)} \right]^{-1} \!\! =
    {\rm i} \eta
    - \epsilon_i - t^2 \! \sum_{\mu \in \partial_i \setminus \nu}\frac{\sum_{l \in \partial_{\mu} \setminus i} \frac{G_l^{(\mu)}}{1+tG_l^{(\mu)}}}{1-t\sum_{l \in \partial_\mu \setminus i} \frac{G_l^{(\mu)}}{1+tG_l^{(\mu)}}} 
    \, ,
\end{equation}
where $\partial_{i}\setminus\nu$ denotes the set of cliques incident to node $i$ except clique $\nu$, and for a given clique 
$\mu$, $l \in \partial_{\mu} \setminus i$ refers to all nodes in $\mu$ except node $i$. A complete derivation of these equations is presented in full detail in Appendix ~\ref{app:cavity}. The Green's function on a given node $i$ can then be expressed in terms of the cavity ones on the neighboring nodes as 
\begin{equation}\label{eq:G}
    G_{ii}^{-1} = {\rm i} \eta 
    - \epsilon_i - t^2 \sum_{\mu \in \partial_i}\frac{\sum_{l \in \partial_\mu \setminus i} \frac{G_l^{(\mu)}}{1+tG_l^{(\mu)}}}{1-t\sum_{l \in \partial_\mu \setminus i} \frac{G_l^{(\mu)}}{1+tG_l^{(\mu)}}} \, .
\end{equation}
Upon averaging over all possible disorder realizations, Eq.~\eqref{eq:cavity} should be interpreted as a self-consistent integral equation for the joint probability distribution of the real and imaginary parts of the cavity Green’s functions [Eq.~\eqref{eq:PG}]. Its stationary solution directly provides the probability distribution of the Green’s functions via Eq.~\eqref{eq:G}, which encodes all relevant observables. These equations can be solved efficiently to arbitrary numerical precision using the population dynamics method~\cite{tikhonov2019critical,rizzo2024localized,tonetti2025testing} (see Appendix ~\ref{app:num} for more details).

\begin{figure*}[t]
    \centering   
    \includegraphics[width=0.5\textwidth]{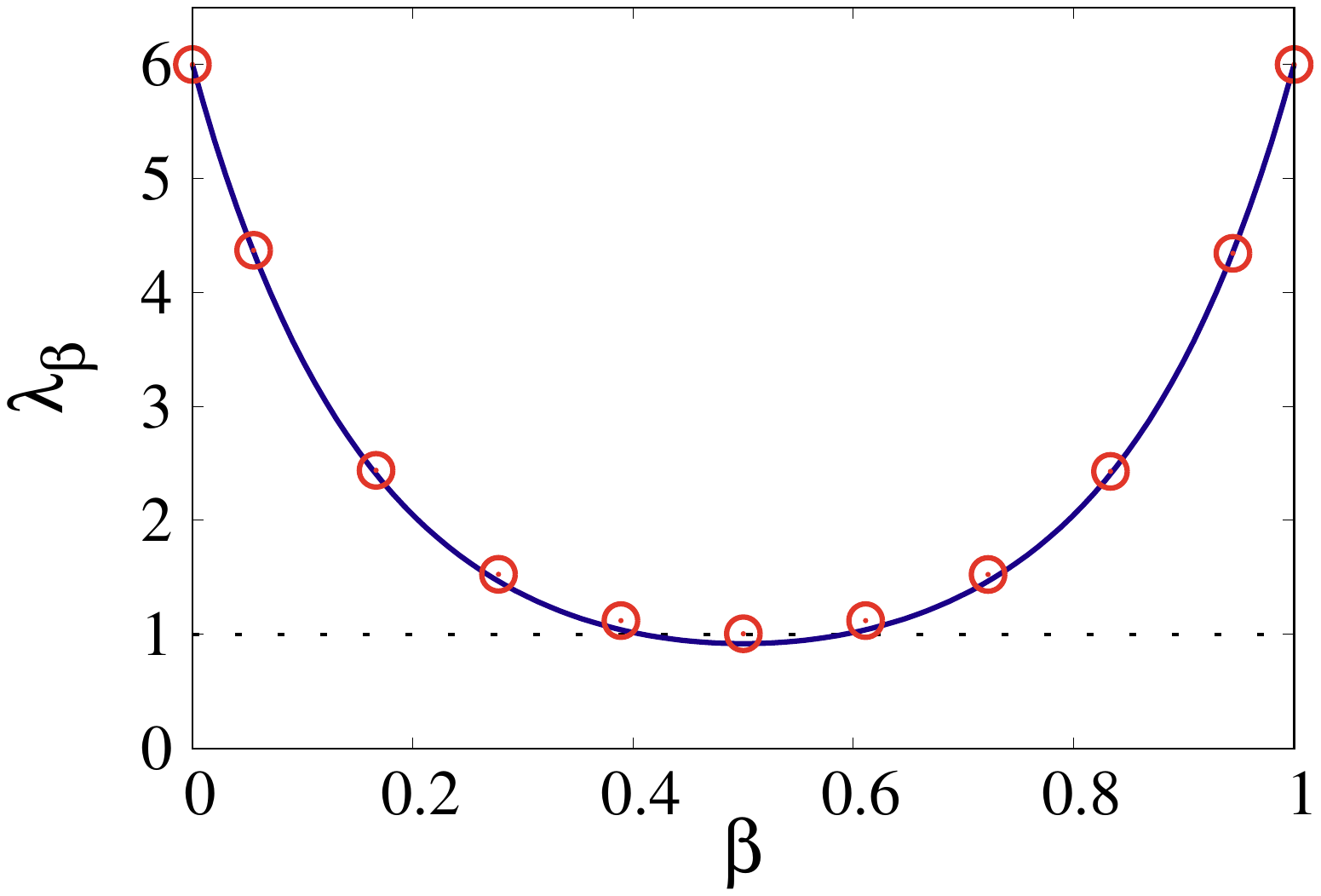} \put(-131,147){(a)}
    \hspace{-0.2cm} \includegraphics[width=0.5\textwidth]{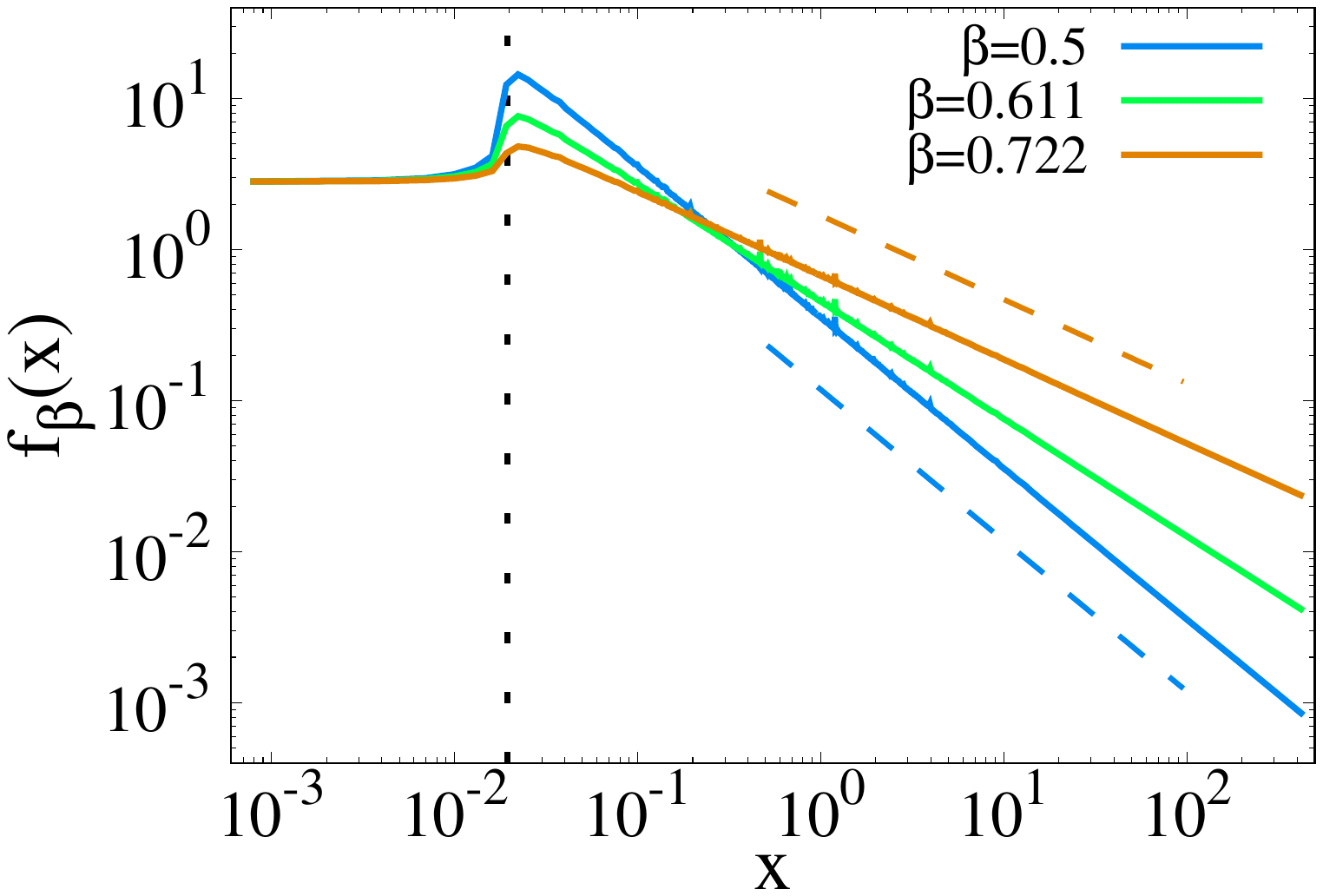}
    \put(-205,147){(b)}
    \vspace{-0.5cm}
    \caption{(a) Largest eigenvalue of the integral operator~\eqref{eq:operator}, obtained from the large-$p$ analytical expression~\eqref{eq:lambda} (blue line) and from direct diagonalization (red circles), for $k=1$, $p=6$, and $W = 103\,t \approx W_c$. 
    (b) Corresponding eigenvectors $f_\beta(x)$ obtained from Arnoldi diagonalization for three values of $\beta$, described by the functional form~\eqref{eq:f}. The vertical dotted line marks $x = 2/W$, and the dashed lines indicate the expected power-laws with exponent $2\beta - 2$. 
    }
    \label{fig:lambda}
\end{figure*}

In the localized phase, the imaginary part of the Green’s function vanishes with the regulator $\eta$. AL can thus be studied from the linear stability of Eq.~\eqref{eq:cavity} with respect to a small imaginary part. To do that, we consider the auxiliary variable $\chi_i^{(\mu)}=\frac{G_i^{(\mu)}}{1+tG_i^{(\mu)}}$, which
leads to a simpler recursion. Writing $\chi_{i}^{(\mu)}=x_{i}^{(\mu)}+i\eta\,\hat x_{i}^{(\mu)}$ and expanding
Eq.~\eqref{eq:cavity} to first order in $\eta$ gives
\begin{eqnarray}\label{eq:Xreal_self} 
    x_i^{(\nu)}&=&\Biggl( t- \epsilon_i - t^2 \!\!\! \sum_{\mu \in \partial_i \setminus \nu} \frac{\sum_{l \in \partial_\mu \setminus i} x_l^{(\mu)}}{1-t\sum_{l \in \partial_\mu \setminus i} x_l^{(\mu)}}\Biggr)^{\! -1} \, , \\
\label{eq:Ximaginary_self} 
    \hat{x}_i^{(\nu)}&=& \left(x_i^{(\nu)}\right)^{\!2} t^2 \!\!\! \sum_{\mu \in \partial_i \setminus \nu} \frac{\sum_{l \in \partial_\mu \setminus i} \hat{x}_l^{(\mu)} }{\left(1-t\sum_{l \in \partial_\mu \setminus i} x_l^{(\mu)} \right)^{2}} \, .
\end{eqnarray}
Equation~(\ref{eq:Xreal_self}) governs the distribution of the real parts $x_{i}^{(\mu)}$ independently of the small imaginary components, while Eq.~(\ref{eq:Ximaginary_self}) is linear in the $\hat{x}_i^{(\mu)}$'s. In the localized regime, the distribution of imaginary components develops a large-$\hat x$ power-law tail, which takes the form~\cite{abou1973selfconsistent,mirlin1991localization,tikhonov2019critical}
\begin{equation} \label{eq:ansatz}
P(x,\hat x)\sim f(x)\,|\hat x|^{-(1+\beta)} \, .
\end{equation}
Inserting this ansatz into the linearized recursion yields a self-consistency condition for $f(x)$ as the eigenvector of unit eigenvalue of a linear integral operator [see Eq.~\eqref{eq:Pxxh}], 
\begin{equation}\label{eq:operator}   
    f(x) = \int d \tilde{x} K_\beta(\tilde{x},x) f(\tilde{x}) \, , 
\end{equation}
where the kernel $K_\beta(\tilde{x},x)$ depends on the distribution of the real part and on the disorder $W$, and is proportional to $kp = z-p$. The explicit form of the kernel is given in Eq.~\eqref{eq:kernel} below. Therefore, a solution of the linearized cavity equations~\eqref{eq:Xreal_self} and \eqref{eq:Ximaginary_self} of the form~\eqref{eq:ansatz} exists only if the largest eigenvalue of this operator is smaller than $1$. This largest eigenvalue can be computed analytically in the limit of large clique connectivity, similarly to the case of the Bethe lattice \cite{abou1973selfconsistent,bapst2014largekAL,tarquini2016level} (see Appendix ~\ref{app:operator} for a fully detailed computation), yielding the asymptotic formula 
\begin{equation} \label{eq:lambda}
    \lambda_\beta (W) \simeq 
    \frac{4 k p t}{W(1 - 2 \beta)} \sinh \! \left[ (2 \beta - 1) \ln \! \left( \frac{2 t}{W}\right) \right] \, .
\end{equation}
Since $\lambda_\beta$ is symmetric around $\beta=1/2$ where it reaches a minimum [see Fig.~\ref{fig:lambda}(a) below], the localization transition occurs at $\beta = \tfrac12$, and the critical disorder $W_c$ is defined by the condition $\lambda_{1/2}(W_c) = 1$, yielding
\begin{equation} \label{eq:Wc}
    \frac{W_c}{t} \simeq 4 k p \ln \left( \frac{W_c}{2 t}\right) 
    \approx 4 (z - p) \ln \left[ 2(z - p) \right] \, .
\end{equation}
To test these analytic predictions, we have computed the largest eigenvalue and its eigenvector using the Arnoldi algorithm~\cite{tikhonov2019critical,parisi2019anderson}. The kernel~\eqref{eq:kernel} is evaluated on a discrete grid with the change of variables $(\tilde x,x)=(\tan\tilde\theta,\tan\theta)$, $\theta\in(-\pi/2,\pi/2)$, with spacing $\Delta\theta=5\times10^{-4}$, and with a uniform discretization of $[-W/2,W/2]$ into $10^3$ points. Results for $k=1$, $p=6$, and $W=103\approx W_c$ are shown in Fig.~\ref{fig:lambda}, demonstrating excellent agreement with Eqs.~\eqref{eq:lambda} and~\eqref{eq:f}. Remarkably, despite relying on a large-$p$ expansion whose leading corrections scale as $1/\ln p$, the accuracy remains high even at moderate $p$.

\begin{figure*}[t]
    \centering   
    \includegraphics[width=0.5\textwidth]{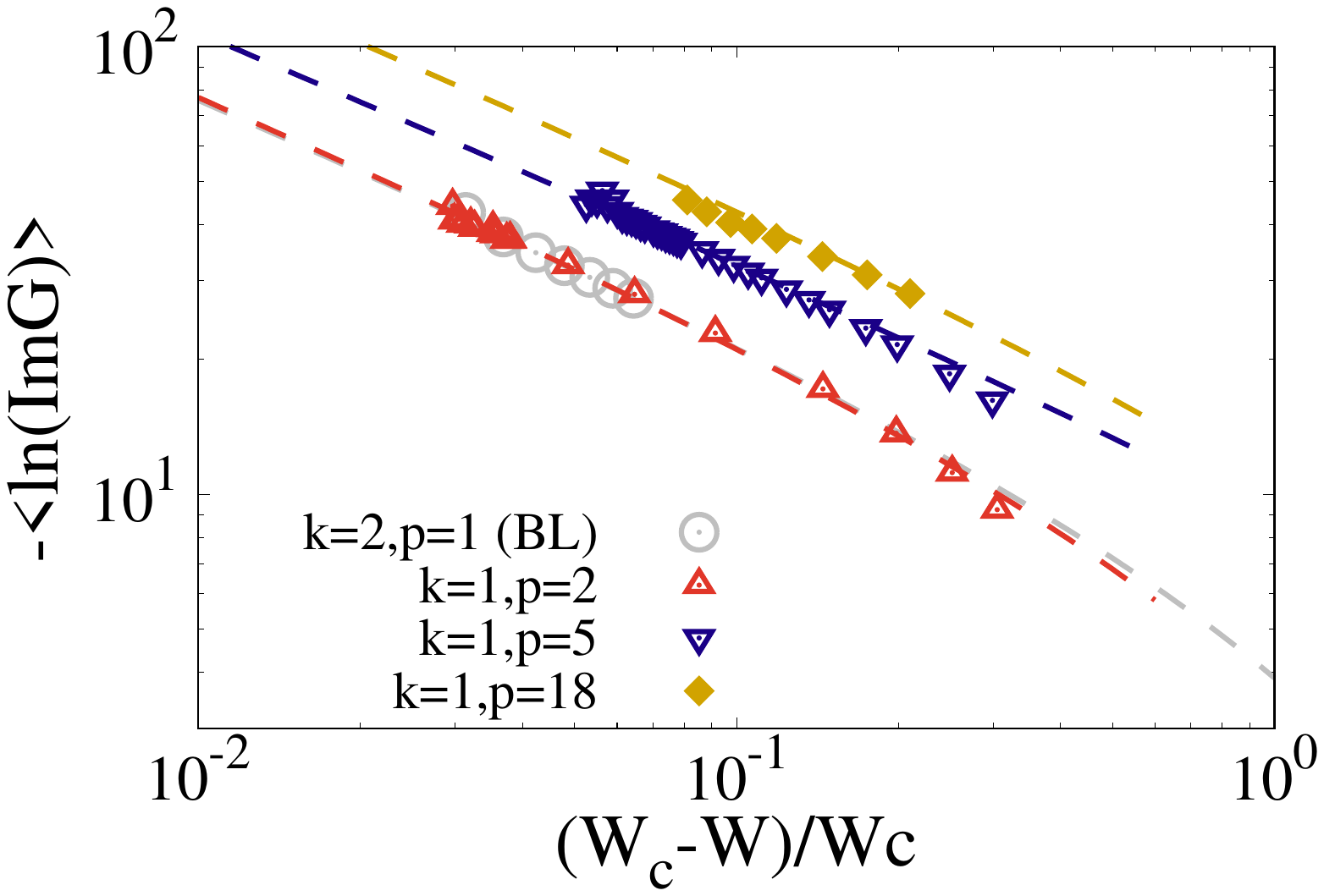} \put(-49,146){(a)}
    \hspace{-0.2cm} \includegraphics[width=0.5\textwidth]{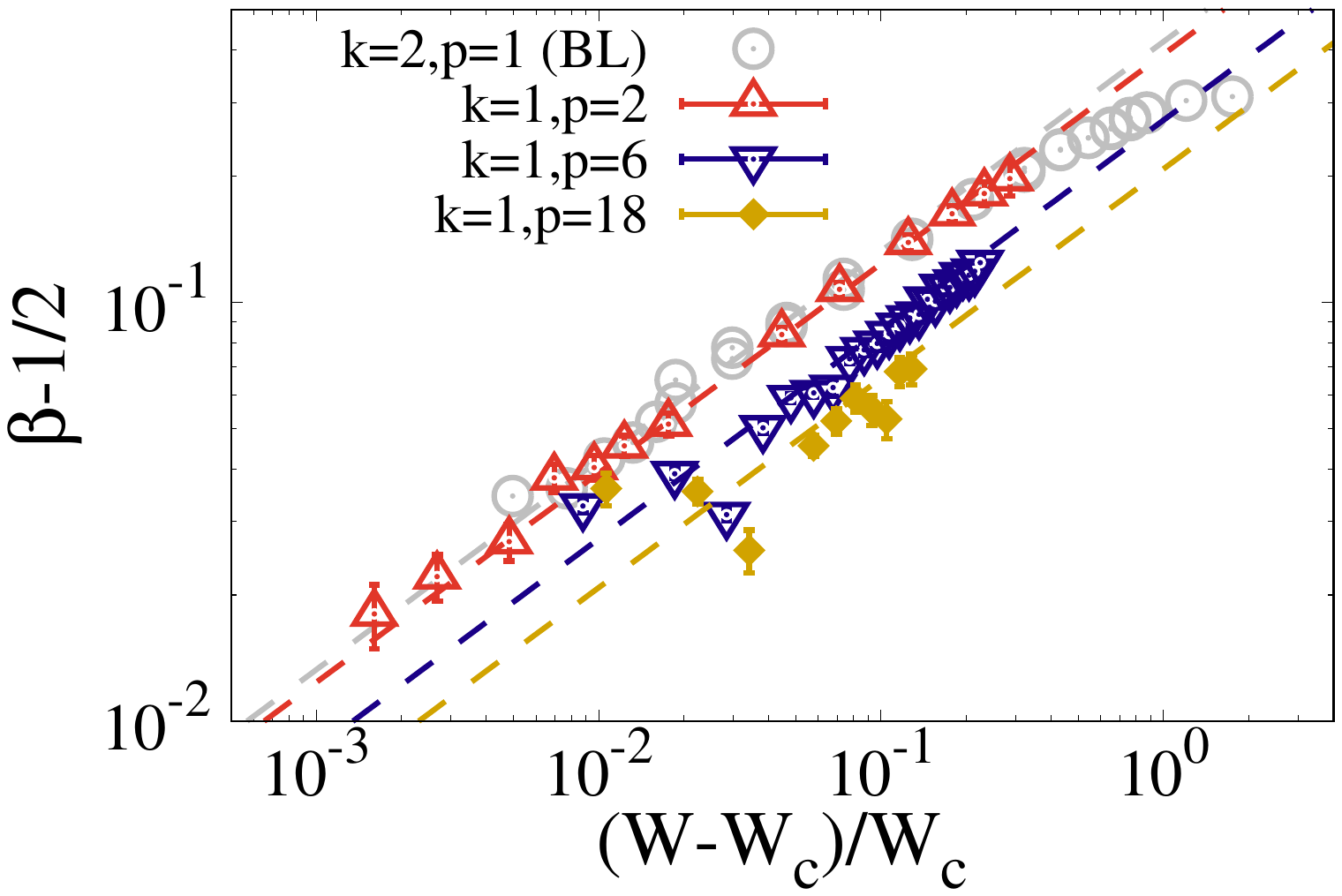} \put(-48,52){(b)} 
    \vspace{-0.7cm}
    \caption{ Critical scaling of the correlation volume, $\ln \Lambda_c \propto -\langle \ln \mathrm{Im}\, G \rangle$ (a), and of the tail exponent $\beta$ (b) of the distribution of the imaginary parts $\hat{x}_i$ [Eq.~\eqref{eq:ansatz}], plotted versus the relative distance from the transition, $|W_c - W|/W_c$, for several pairs $(k,p)$ controlling the graph topology. Symbols correspond to results obtained by solving the cavity equations using population dynamics with large populations of size $\Omega = 2^{28} \simeq 2.7 \times 10^{8}$ (see Appendix ~\ref{app:num} for details). Dashed curves show the predictions of Eqs.~\eqref{eq:Lambdac} and~\eqref{eq:beta}, with coefficients $c_1$ and $c_2$ from Eq.~\eqref{eq:lambda_exp}.}
    \label{fig:critical}
\end{figure*}

\subsection{Critical behavior}
The critical behavior is fully encoded in the dependence of $\lambda_\beta(W)$ close to the critical point~\cite{zirnbauer1986anderson,zirnbauer1986localization,verbaarschot1988graded,tikhonov2019critical,tikhonov2019statistics,mirlin1991localization,mirlin1994statistical}. Expanding $\lambda_\beta$ near $W_c$ and $\beta=1/2$ gives
\begin{equation} \label{eq:lambda_exp}
\begin{aligned}
    & \lambda_{\beta} (W)  \simeq 1 - c_1 (W - W_c) + c_2 \left( \beta - \frac{1}{2} \right)^2 \, , \\
    &c_1  = \frac{t}{W_c} \left[ 1 - \ln^{-1} \left( \frac{W_c}{2 t} \right) \right ] \, , 
    \qquad c_2 
    = \frac{2}{3} \ln \left( \frac{W_c}{2 t} \right) \, ,
    \end{aligned}
\end{equation}
with $W_c$ fixed by $\lambda_{1/2}(W_c)=1$, yielding Eq.~\eqref{eq:Wc}. As shown in Appendix ~\ref{app:corr} in full detail, in the localized phase the two-point function that describes the spatial correlations between eigenstates' amplitudes is asymptotically governed at large distances by the same integral operator~\eqref{eq:operator} that controls the stability of the imaginary part of the Green’s function under iteration in the linearized regime. This implies that approaching the critical point from the localized phase the localization length diverges as [see in particular Eqs.~\eqref{eq:C0r1} and \eqref{eq:C0r2} of Appendix ~\ref{app:corr}]
\begin{equation}\label{eq:loc_len}
\xi_{\rm loc} \simeq -[\ln \lambda_{1/2}]^{-1} \simeq [c_1 (W - W_c)]^{-1} \, .
\end{equation} 
The exponent of the tails of the distribution of the imaginary parts $\hat{x}_i$ in Eq.~\eqref{eq:ansatz} is fixed by the condition $\lambda_\beta(W) = 1$ which, for $W \gtrsim W_c$ gives~\cite{rizzo2024localized}
\begin{equation} \label{eq:beta}
    \beta \simeq \frac{1}{2} + 
\sqrt{\frac{c_1}{c_2}} \sqrt{W - W_c} \, .
\end{equation} 
In the delocalized phase, but still close to the transition ($W \lesssim W_c$), the linearization~\eqref{eq:Xreal_self} and \eqref{eq:Ximaginary_self} ceases to be justified, although the imaginary parts of the Green’s functions remain small on most sites. In this regime, the operator~\eqref{eq:operator} captures the characteristic scale, known as the correlation volume $\Lambda_c$, beyond which the linear approximation breaks down~\cite{mirlin1994distribution,mirlin1991localization,mirlin1994statistical,biroli2022critical}. The asymptotic behavior of $\Lambda_c$ is obtained from the imaginary part of the solution of $\lambda_\beta(W) = 1$ for $W \lesssim W_c$ when $\beta$ is analytically continued to the complex plane \cite{mirlin1991localization,mirlin1994statistical,tikhonov2019critical} yielding

\begin{equation} \label{eq:Lambdac}
\ln \Lambda_c \propto - \langle \ln {\rm Im} G \rangle \simeq \frac{\pi \sqrt{c_2/c_1}}{\sqrt{W_c - W}} \, .
\end{equation}
To test these asymptotic analytic predictions for the critical behavior, we performed extensive population dynamics calculations to solve the cavity equations numerically close to the critical point, using very large population sizes (see Appendix ~\ref{app:num} for details). Specifically, Fig.~\ref{fig:critical}(a) shows the critical behavior of the correlation volume $\Lambda_c$ as $W_c$ is approached from the metallic side of the transition, while Fig.~\ref{fig:critical}(b) shows the critical behavior of the exponent $\beta$ governing the tails of the probability distribution of the imaginary parts $\hat{x}_i$ in Eq.~\eqref{eq:ansatz} when the transition is approached from the insulating phase.
In both cases, we find excellent agreement between the population dynamics results (symbols) and the analytic predictions of Eqs.~\eqref{eq:beta} and \eqref{eq:Lambdac} (dashed curves), where the coefficients $c_1$ and $c_2$ are not treated as fitting parameters but are fixed by Eqs.~\eqref{eq:lambda_exp}.

Note that the peculiar form of criticality described here is identical to that observed on the loopless Bethe lattice, indicating that the presence of a finite density of local loops of arbitrary (but finite) sizes does not modify the critical behavior~\cite{baroni2024corrections}.

\begin{figure*}[t]
    \centering
    \includegraphics[width=0.42\textwidth]{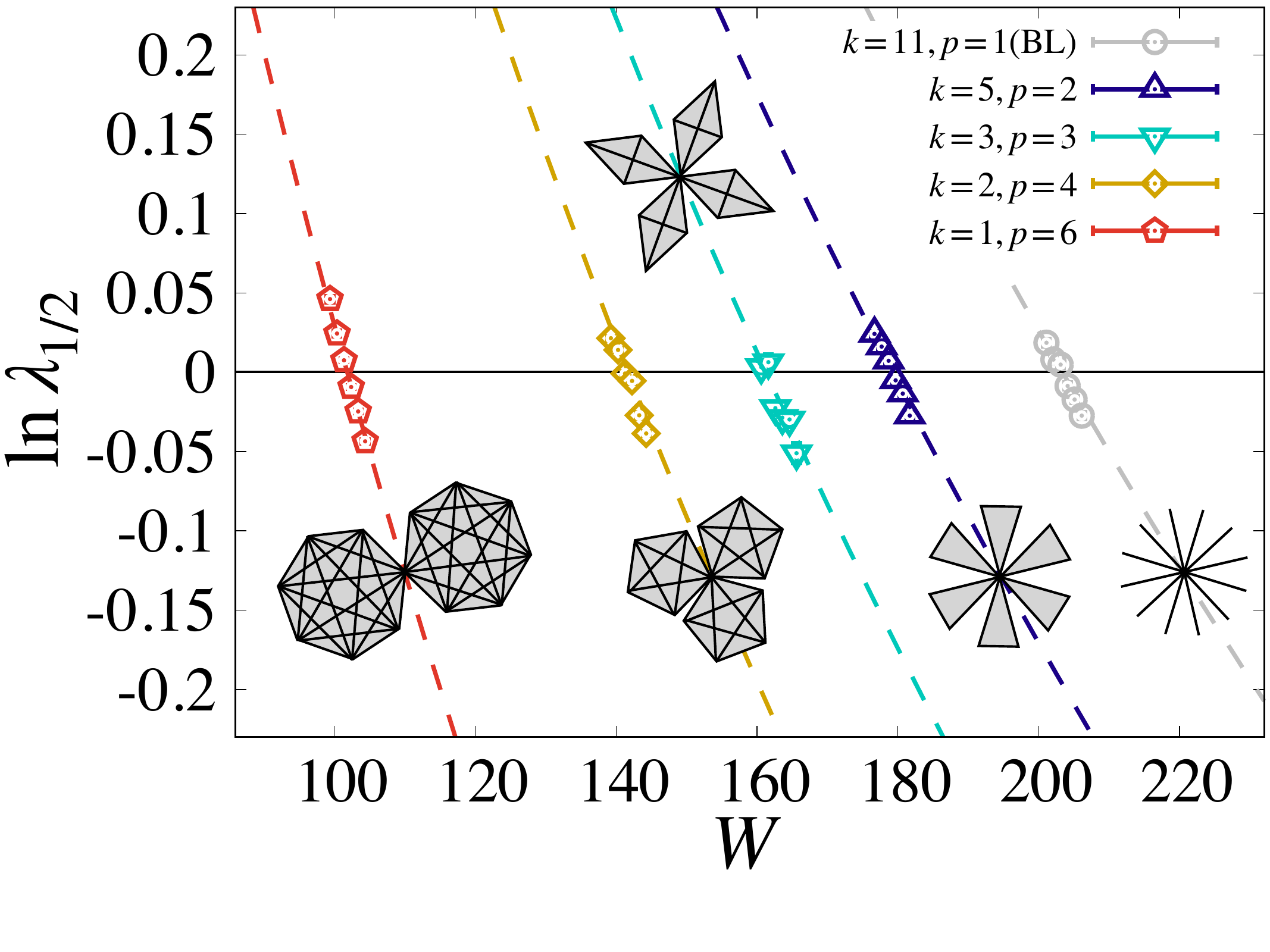} \put(-155,149){(a)}
    \hspace{1.cm}
    \includegraphics[width=0.42\textwidth]{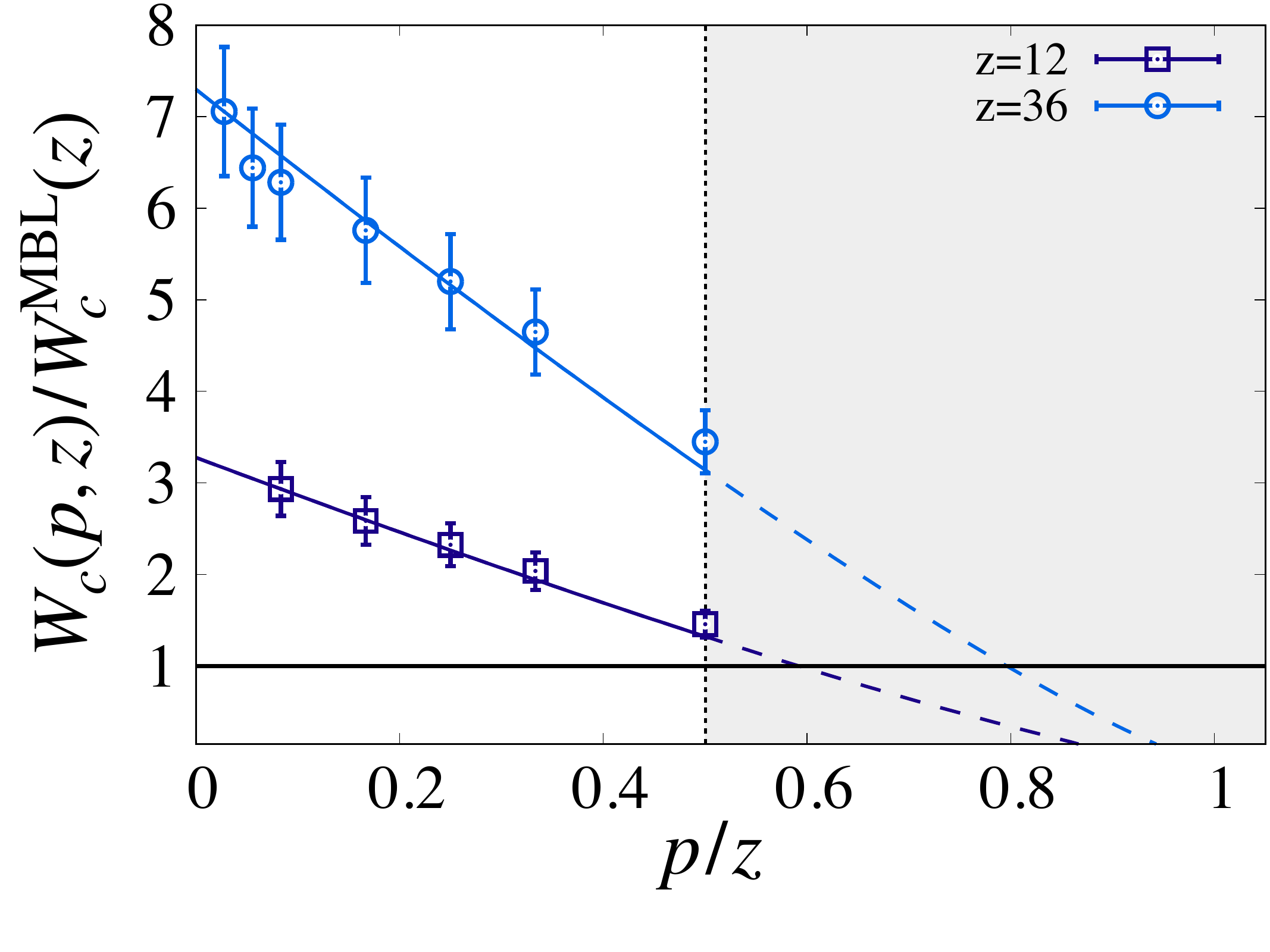} \put(-160,146){(b)}
    \vspace{-0.6cm}
    \caption{(a) Logarithm of the exponent, $\ln \lambda_{1/2}$, controlling the exponential evolution of the typical imaginary part of the cavity Green’s functions~\eqref{eq:Ximaginary_self} under iteration, plotted as a function of the disorder strength $W$. Each data set corresponds to a different loop topology $(k,p)$  at fixed connectivity $z=12$. The dashed lines represent the analytical prediction of Eq.~\eqref{eq:lambda}, while symbols show the population dynamics results with large populations of size $\Omega = 2^{28} \simeq 2.7 \times 10^{8}$ (see Appendix ~\ref{app:num} for details).
    (b) Ratio between the critical disorder of the Anderson model on the Husimi tree, $W_c(p,z)$, varying $k$ and $p$ at fixed $z$, and the effective diagonal disorder strength of an interacting disordered spin chain~\cite{imbrie2016diagonalization,de2024absence,biroli_large-deviation_2024},
    $W_c^{\rm MBL}(n)= 2\sqrt{n}, h_c$, with $h_c \approx 9$ the estimated MBL transition~\cite{biroli_large-deviation_2024}, and $n=z$.
    Solid lines show the analytical predictions~\eqref{eq:Wc} for $p/z \le 1/2$ for $z=12$ and $z=36$, while dashed lines indicate their analytic continuation into the loop-rich regime relevant for MBL ($p/z > 1/2$, shaded area). Symbols (squares for $z=12$ and circles for $z=36$) denote the critical disorder estimated from the zero crossing of a linear fit to the population dynamics data for $\ln \lambda_{1/2}$.}
    \label{fig:W_c}
\end{figure*}

\section{The critical disorder}
As explained above, the largest eigenvalue $\lambda_{1/2}(W)$ controls the exponential growth (for $W < W_c$) or decay (for $W > W_c$) of the typical value of the $\hat{x}_i^{(\mu)}$'s under iteration, via $\ln \lambda_{1/2} = \lim_{n_{\rm iter} \to \infty} \langle \ln \hat{x} \rangle / n_{\rm iter}$. Figure ~\ref{fig:W_c}(a) displays $\ln \lambda_{1/2}$ as a function of $W$ at fixed connectivity $z = 12$, varying the graph topology through the parameters $(k,p)$. The dashed curves give the analytical expressions of Eq.~\eqref{eq:lambda}, showing excellent agreement with the population dynamics data (symbols, see Appendix ~\ref{app:num} for details), even for moderate clique sizes. 

The crucial observation is that the critical disorder $W_c$ systematically decreases with increasing clique size $p$, demonstrating that local loops enhance localization. Intuitively, loops create multiple interfering return paths, effectively reducing the number of independent directions through which amplitude can propagate. This mechanism is quantified by our analytic result: the eigenvalue ~\eqref{eq:lambda} has the same form as for the loopless Bethe lattice~\cite{abou1973selfconsistent,tikhonov2019critical,mirlin1991localization}, but with the branching ratio $z-1$ replaced by $k p$, which is strictly smaller than the local degree. The factor $k p$ can be in fact interpreted as the effective branching ratio controlling the exponential growth of the number of nodes at distance $r$, which on the Husimi tree scales as $(k p)^r = (z-p)^r$,  which is significantly below the Bethe lattice scaling $(z-1)^r$ for the same connectivity, reflecting the overcounting of paths induced by loops.

A complementary viewpoint comes from the linearized recursion~\eqref{eq:Ximaginary_self}, which differs from the Bethe lattice equation~\cite{abou1973selfconsistent,rizzo2024localized} only through the renormalized hopping $t_{\mu\to 0}^{\rm eff} = t / \bigl( 1 - t \sum_{l \in \partial_\mu \setminus 0} x_l^{(\mu)} \bigr)$. For typical sites where $x_l^{(\mu)}$ is small, $t^{\rm eff}\!\approx t$. However, if a neighbor is resonant (i.e. with a large $x_l^{(\mu)}$), the denominator becomes large and $t^{\rm eff}$ is strongly suppressed, effectively isolating the entire clique and acting as a transport blockade. Larger cliques have a higher probability of hosting such resonances and a wider impact when they occur, naturally leading to the observed reduction of $W_c$ with increasing $p$.

\subsection{Implications for MBL} \label{sec:MBL}
As a representative example, consider the random-field Ising chain of $n$ spins in a transverse field, for which absence of standard diffusion at strong disorder has been rigorously established~\cite{imbrie2016diagonalization,de2024absence,biroli_large-deviation_2024}. Its Hilbert-space graph is the $n$-dimensional hypercube, with degree $z=n$. Since the on-site energies are sums of $n$ random fields $h_i\in[-h,h]$, the effective strength of the diagonal disorder at the MBL transition scales as $W_c^{\rm MBL}(n)\approx 2\sqrt{n}\,h_c$.  To connect with our single-particle model on the Husimi tree, Fig.~\ref{fig:W_c}(b) shows the ratio $W_c(p,z)/W_c^{\rm MBL}(z)$ versus $p/z$ for $z=12$ and $36$. The large-$p$ prediction~\eqref{eq:Wc} (solid lines) agrees well with the numerical values extracted from the vanishing of $\ln\lambda_{1/2}$ (panel (a)). The dashed curves represent the analytic continuation of Eq.~\eqref{eq:Wc} to $p>z/2$, outside the domain of model~\eqref{eq:hamiltonian}. This continuation indicates that, for any fixed $z$, increasing $p$ eventually reaches a regime $p\lesssim z$ where the Husimi-tree localization threshold matches the scaling of $W_c^{\rm MBL}$.

Hilbert-space graphs contain far more loops than the Husimi tree. On the hypercube, loops range from length $4$ up to $2^n$, and permutations of spin-flip order on closed cycles yield an approximate count of the number of loops of length $\ell$ as ${\cal N}_{\rm HC}(\ell)\simeq {\cal N} (\ell) \simeq \tfrac 1 2 n^{\frac{\ell}{ 2}} \tfrac \ell 2 !$ (see Appendix ~\ref{app:loops} for a detailed explanation). Interpreting $p$ as a proxy for the typical loop size $\langle\ell\rangle$, the extrapolation in Fig.~\ref{fig:W_c}(b) suggests that sufficiently loop-rich graphs naturally push $W_c$ toward the MBL scaling.

A complementary perspective uses the notion of an effective branching ratio. On the hypercube, the number of configurations at Hamming distance $r$ is $\binom{n}{r}$, defining an $r$-dependent effective branching ratio $(kp)_{\rm eff}=[\binom{n}{r}]^{1/r}$, which decreases from $n$ at $r=1$ to $2$ at $r=n/2$. Via Eq.~\eqref{eq:Wc}, this implies that as $r$ grows, one eventually reaches $r\sim\sqrt{n}$ where the effective branching ratio becomes small enough to allow localization at $W_c^{\rm MBL}\simeq 2\sqrt{n}\,h_c$.

\begin{figure*}[t]
    \centering
    \includegraphics[width=0.42\textwidth]{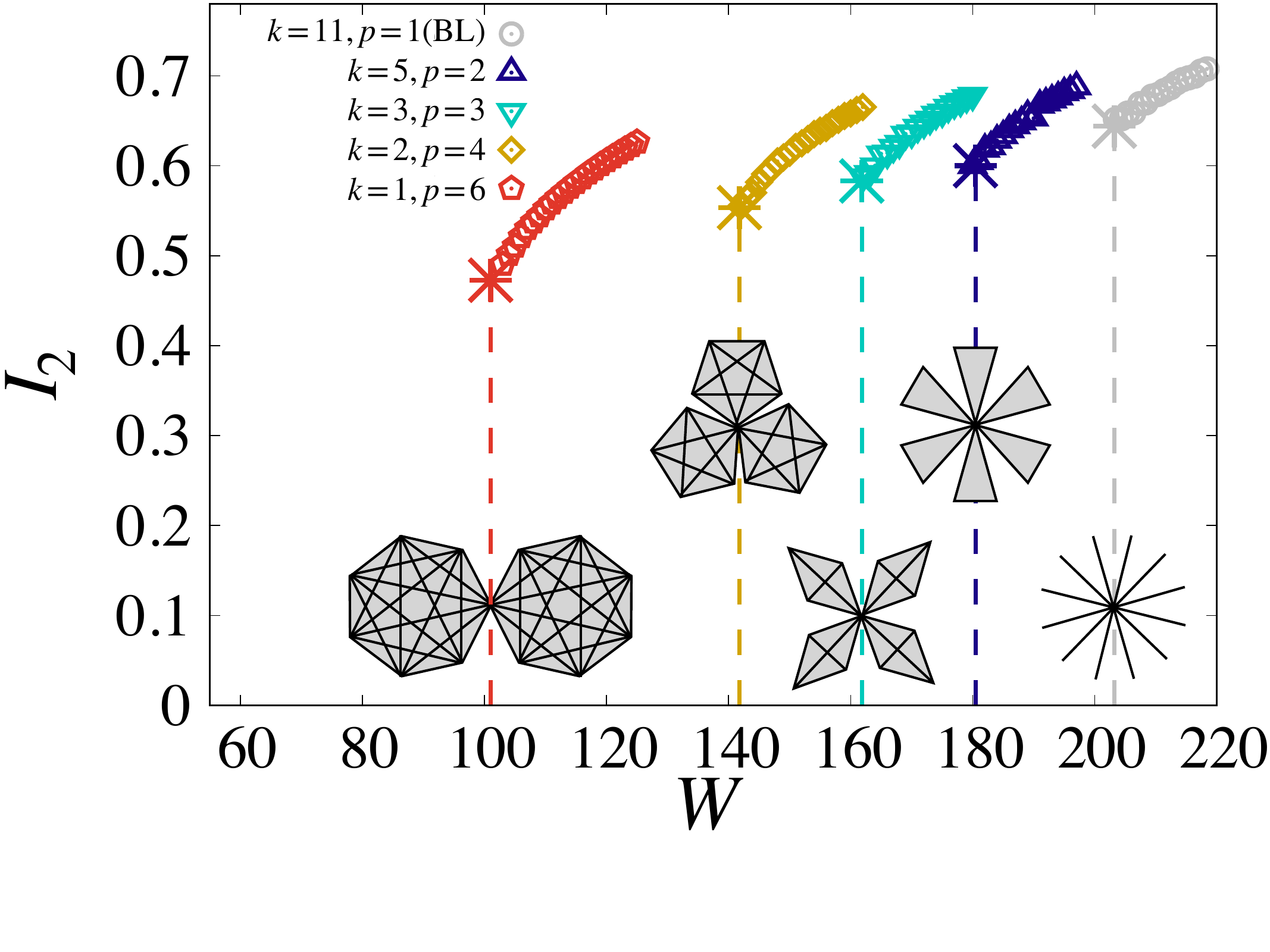} \put(-108,150){(a)}
    \hspace{0.8cm}
    \includegraphics[width=0.42\textwidth]{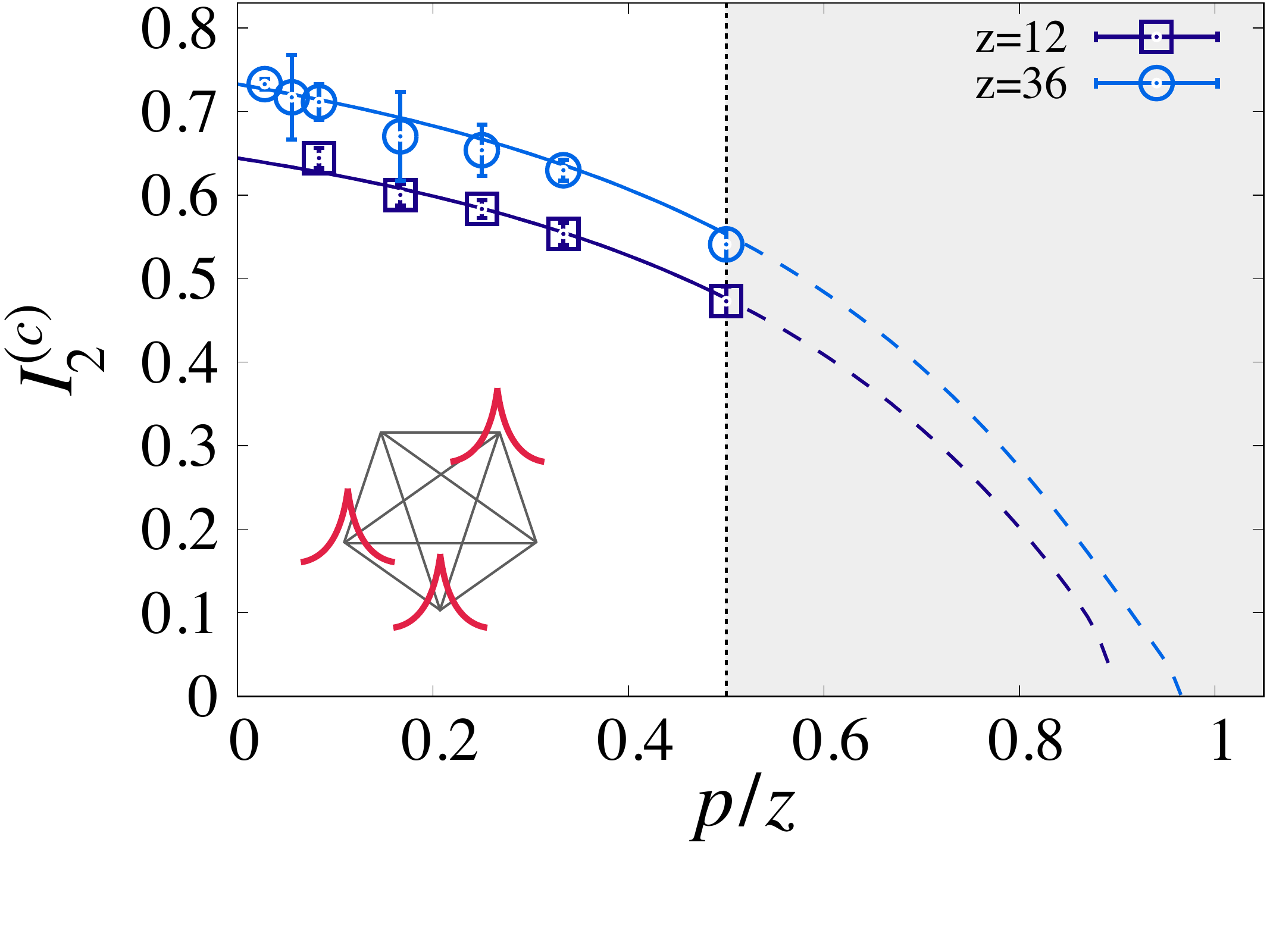} \put(-114,149){(b)}
    \vspace{-0.8cm}
    \caption{(a) IPR as a function of disorder strength $W$ for Husimi trees with fixed connectivity $z=12$ varying $k$ and $p$. The population dynamics results are obtained using the procedure described in Appendix ~\ref{app:num_IPR} with large populations of size $\Omega = 2^{28} \simeq 2.7 \times 10^{8}$.
    (b) Magnitude of the IPR jump at the critical threshold, $I_2^{(c)}$, as a function of the loop-to-connectivity ratio $p/z$ for different connectivities $z=12$ (squares) an $z=36$ (circles). Solid lines are fits to the toy model of Eq.~\eqref{eq:toyIPR}, using a single parameter $\gamma \simeq 5$ for both curves. The analytic continuation into the MBL-relevant regime (dashed lines, shaded region) shows that the jump can be suppressed to arbitrary small values for $p \lesssim z$. }
    \label{fig:IPR}
\end{figure*}

\section{Inverse Participation Ratio}
We now turn to the second key difference between single-particle AL on hierarchical graphs and MBL, namely the scaling of the IPR in the localized phase. The IPR can be efficiently computed from the numerical solution of the exact recursion relations~\cite{rizzo2024localized,tonetti2025testing} (see Appendix ~\ref{app:num_IPR} for a detailed explanation). Figure ~\ref{fig:IPR}(a) shows $I_2$ as a function of disorder $W$ for fixed connectivity $z=12$ and varying clique size $p$. The plot clearly reveals that the IPR decreases systematically with increasing $p$, indicating that the localized states become more extended. 

This effect can be captured quantitatively by a simple first-order perturbative argument. Starting from a localized state of zero energy on a given node, the average number of resonant nodes within its clique at the critical disorder is proportional to $t/W_c$. Assuming that the wave function spreads uniformly over all resonant peaks [as depicted in the sketch in the inset of Fig.~\ref{fig:IPR}(b)], the corresponding reduction of the jump of the IPR at the localization transition can be expressed as
\begin{equation}\label{eq:toyIPR}
I_2^{(c)}(z,p) = \frac{I_2^{(c,BL)}(z)}{1 + \gamma \frac{t p}{W_c(z,p)}} \, ,
\end{equation}
where $\gamma$ is an adjustable constant of order unity. Fig.~\ref{fig:IPR}(b) compares this simple expression (solid lines) to the numerical data for $I_2^{(c)}$ at $z=12$ and $36$, using a common value of $\gamma \simeq 5$. The agreement is remarkably good given the simplicity of the approximation. Extending this model beyond the parameter range of the Husimi tree, into the loop-rich regime $p/z > 1/2$ (shaded region), shows that the IPR jump can be made arbitrarily small (dashed lines), consistent with the expected behavior of MBL eigenstates.

Equation~\eqref{eq:toyIPR} also clarifies the mechanisms behind the reduced IPR: (i) loops lower the critical disorder $W_c$, increasing the probability of finding nearby resonances, and (ii) fully connected cliques amplify the effect of these resonances, promoting the spreading the localized wave function.

\section{Conclusions}
In this paper we have studied Anderson localization on a class of solvable models that feature a finite density of local loops of arbitrary length. This offers a minimal framework to study the impact of loops on localization, and demonstrates that local loops, a key topological feature of many-body Hilbert-space graphs, fundamentally impact the AL transition in a way that resolves central discrepancies with MBL phenomenology. On the one hand, local loops reduce the graph's effective branching ratio, thereby favoring localization and systematically lowering the critical disorder $W_c$; On the other hand, the loop structure enhances local hybridization, increasing the spatial extent of eigenstates and therefore decreasing the IPR of localized states. The analytical solutions of the model offer a clear understanding of the physical mechanisms behind these phenomena. Crucially, the universality class remains that of the loopless Bethe lattice~\cite{garcia2022critical}. 

Yet, the loop structure of realistic many-body Hilbert-space graphs is significantly more complex than that of Husimi trees, and the influence of loops cannot be encoded in a single parameter $p$. Moreover, loops are only part of the story: correlations among matrix elements play an equally crucial role~\cite{roy2020fock, roy2020localization, scoquart2024role}. The example of the QREM illustrates this clearly---the loops of the hypercube alone are not sufficient to produce MBL at finite disorder (in the middle of the spectrum) when on-site energies are uncorrelated random variables~\cite{Faoro_2019,baldwin2018quantum,biroli2021out,parolini2020multifractal,Smelyanskiy_2020}. Nevertheless, Husimi trees emerge as the minimal solvable framework capturing both the local-loop structure and Anderson criticality, opening a controlled path toward a realistic theory of the MBL transition. Another fundamental example of the subtle and crucial role of correlations concerns the fate of MBL in higher-dimensional systems. In fact, the Hilbert-space graph of a disordered Ising model in a transverse field on a $d$-dimensional Euclidean lattice of $n$ spins is a hypercube with local degree $n$, independently of $d$. Yet it is generally accepted that the MBL phase is eventually destroyed for $d>1$ in the infinite-size, infinite-time limit by rare thermal regions where the disorder is anomalously weak~\cite{deroeck2017stability}. From the Hilbert-space perspective discussed in Sec.~\ref{sec:MBL}, the distinction between $d=1$ and $d>1$ manifests only through subtle differences in the statistics of the random energies and in their correlation structure, which inherit the geometry of the underlying lattice through the spatial dependency  of the antiferromagnetic interacting term of the many-body Hamiltonian. Nevertheless, despite these caveats, Husimi trees emerge as the minimal analytically tractable framework that simultaneously captures local loop structure and Anderson criticality, thereby opening a controlled path toward a more realistic theory of the MBL transition.

This framework is naturally extensible, opening several promising directions for future research. A natural next step is to study Husimi trees with random cliques' shapes and/or to incorporate (short-range) correlations in the random energies and local hopping terms~\cite{roy2020fock, roy2020localization, scoquart2024role}.
Another important avenue is the development of a fully analytical treatment for the IPR near the critical point, going beyond the simple toy model~\eqref{eq:toyIPR}. Finally, generalizing the model to graphs without the $p/z \le 1/2$ constraint~\cite{biroli2021out,toninelli2005cooperative} would remove the need for extrapolation and provide a definitive characterization of the phase diagram for loop-rich graphs.

\section{Acknowledgments}
We are grateful to G. Biroli, N. Laflorencie, G. Lemari\'e, A.D. Mirlin, T. Scoquart, D. Venturelli for insightful comments and discussions. We acknowledge financial support from the ANR research grant ManyBodyNet ANR-24-CE30-5851. 

\appendix

\section{Loop statistics: Husimi trees versus Hilbert space graphs of interacting systems}\label{app:loops}

In a Husimi tree, loops are limited to occur within the $(p+1)$-cliques, since any two cliques intersect in a single node. This prevents loops from propagating across the graph and restricts their lengths to the interval $3 \le \ell \le p+1$. The number of loops of length $\ell$ originating from any given node of the Hisimi tree is
\begin{equation}\label{eq:loops}
    \mathcal{N} (\ell)=(k+1) \binom{p}{\ell-1}\frac{(\ell-1)!}{2}= \frac{(k+1) \, p! }{2 \, (p - \ell + 1)!} \, .
\end{equation}
The meaning of each factor in this expression is straightforward. The prefactor $(k+1)$ counts the number of cliques to which the loop can belong. The binomial coefficient gives the number of ways to choose the $\ell-1$ nodes among the other $p$ nodes of the clique (besides the origin) that form the loop. The term $(\ell-1)!$ accounts for all possible permutations of the order in which these nodes are visited along the loop. Finally, the factor $1/2$ prevents double counting of the same loop when traversed in opposite directions.

\begin{figure*}[]
    \centering
    \includegraphics[width=0.42\textwidth]{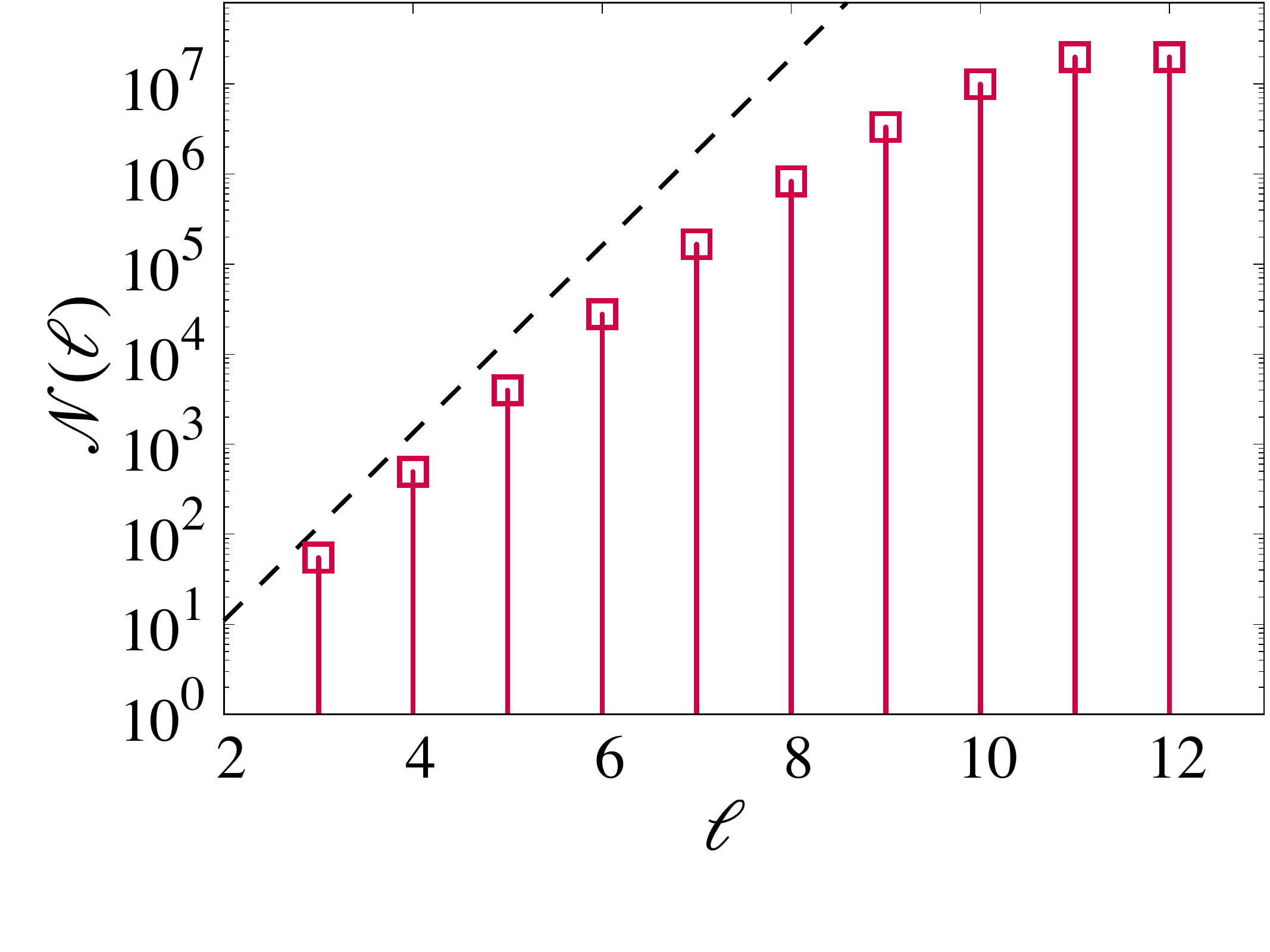} \put(-173,150){(a)}
    \hspace{0.8cm}
    \includegraphics[width=0.42\textwidth]{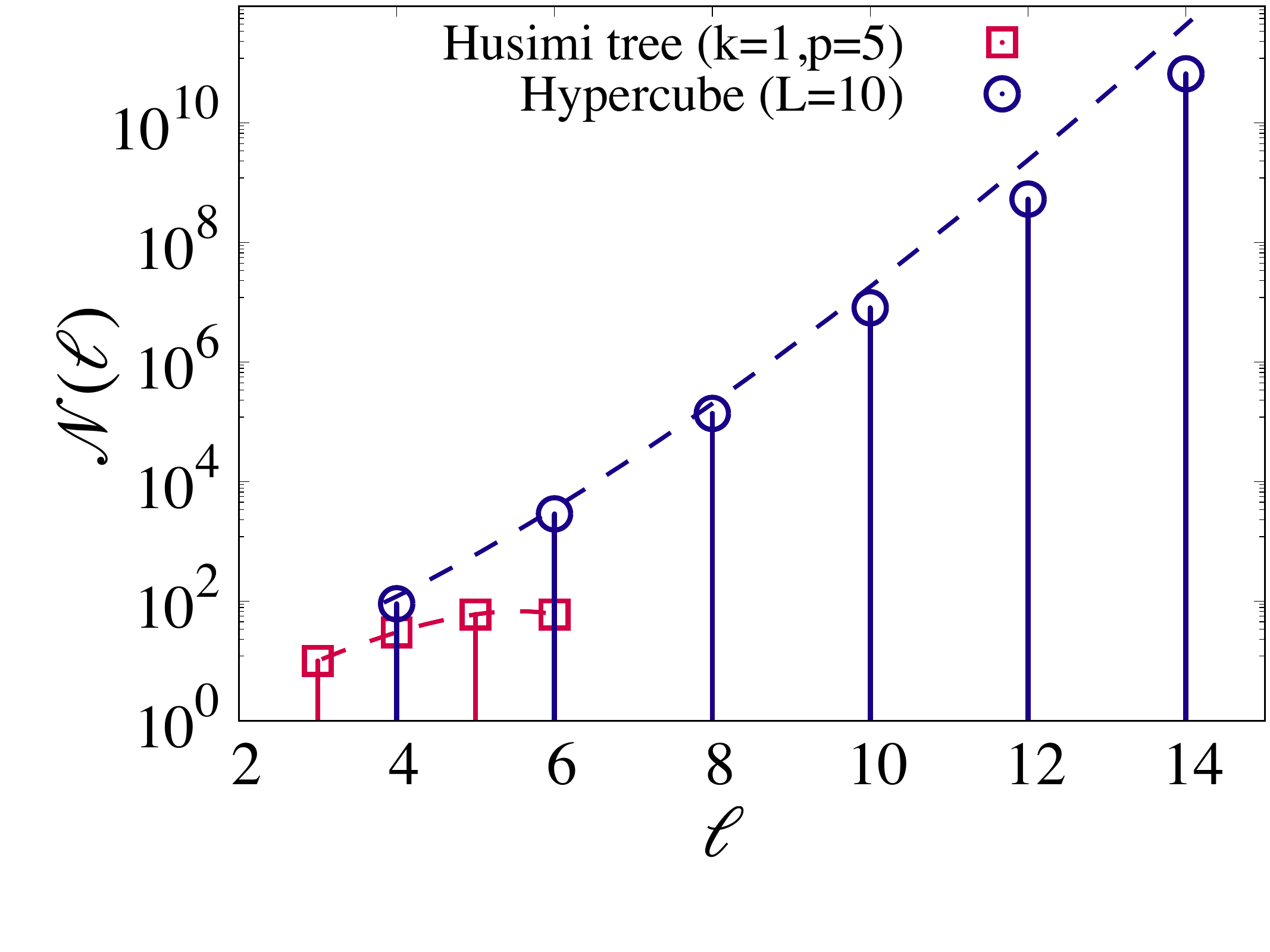} \put(-170,149){(b)}
    \vspace{-0.8cm}
    \caption{(a) Number of loops of length $\ell$ starting from a given node of a Husimi tree with $k=1$ and $p=11$, as given by Eq.~\eqref{eq:loops}. The dashed line shows the approximate exponential scaling ${\cal N}(\ell) \approx (k+1)p^{\ell}/2$. (b) Comparison of loop statistics for the Husimi tree and the hypercube with the same local connectivity $z=10$. Blue circles denote the number of loops of length $\ell$ starting from a given vertex of the hypercube formed by the $2^{10}$ bit strings $\{+1,-1\}^{10}$, with $\ell \ge 4$ and even, obtained via an exact counting algorithm. The dashed blue line corresponds to the asymptotic expression~\eqref{eq:loopsHC}, valid for large $n$ and $\ell \lesssim n$. Red squares denote the number of loops of length $\ell$ in a Husimi tree with $k=1$ and $p=5$.}
    \label{fig:loops}
\end{figure*}

The number of loops starting from a given node of the Husimi tree is plotted in Fig.~\ref{fig:loops}(a) for $k=1$ and $p=11$. It is easy to check from Eq.~\eqref{eq:loops} that $\mathcal{N} (\ell)$ grows roughly exponentially with $p$, as ${\cal N}(\ell) \approx (k+1)\, p^{\ell}/2$. As a consequence, the statistics of loops is dominated by the longest allowed loops, whose typical size is of order $p$. Hence, on the Husimi tree, the parameter $p$ controls both the number and the typical length of loops, and thus fully characterizes the loop statistics.

However, the behavior described by Eq.~\eqref{eq:loops} is markedly different from that of generic Hilbert-space graphs of many-body systems. As a reference example, let us consider the $n$-dimensional hypercube with $2^n$ nodes, corresponding to the bit-strings $\{+1,-1\}^n$ of length $n$, which represents the Hilbert-space graph of a generic quantum spin chain in the basis of simultaneous eigenstates of the $\{\hat{\sigma}_i^z\}$ operators.  On this graph, loops (ov even length) range from length $4$ up to length $2^n$.  One can obtain an asymptotic estimate of the number of loops of length $\ell \le n$ starting from a given node (i.e., a given spin configuration) as follows.  Consider directed paths of length $\ell$ constructed by flipping $\ell/2$ spins in some order, and then flipping the same spins back in a different order.  We can approximately count the number of such cycles.  First, there is a binomial factor $\binom{n}{\ell/2}$ accounting for the choice of which $\ell/2$ spins out of $n$ are flipped.  Next, there is a factor $(\tfrac\ell2!)^2$ corresponding to the permutations of the flip order in the forward and backward halves of the cycle.  (Note that one should exclude from this counting the paths that cross. However, since the hypercube is effectively $n$-dimensional, in the large-$n$ limit the probability that two random permutations of the forward and backward spin-flip order intersect at one (or more) given spin configuration(s) along the path vanishes as $n \to \infty$.) Finally, a factor $1/2$ avoids double counting of the same loop traversed in opposite directions.  Of course, these directed cycles are not the only loops of length $\ell$ (more general, nondirected cycles also exist), but they constitute the dominant contribution for relatively short loops $\ell \lesssim n$. Putting all pieces together one obtains
\begin{equation} \label{eq:loopsHC}
    \begin{aligned}
        {\cal N}_{\rm HC}(\ell) &\gtrsim 
    \frac{1}{2}\binom{n}{\ell/2} \left(\tfrac\ell2!\right)^2\\
    &= 
    \frac{n (n-1)\cdots(n-\ell/2 + 1)}{2}\, \frac{\ell}{2}!
    \approx 
    \frac{n^{\ell/2}}{2} \, \frac{\ell}{2}! \, ,
    \end{aligned}
\end{equation}
where the last expression holds asymptotically for large $n$ and $\ell \lesssim n$.

In Fig.~\ref{fig:loops}(b) we compare the number of loops originating from a given node in two cases: a Husimi tree with $k=1$ and $p=5$, and a hypercube corresponding to the Hilbert-space graph of a spin chain of length $n=10$ (computed numerically using an exact counting algorithm).  The parameters are chosen such that both graphs have the same local degree $z=10$.  The plot clearly shows that the hypercube possesses vastly more loops---and loops of much larger lengths---than the Husimi tree, reflecting the far richer loop structure of generic many-body Hilbert-space graphs.

\section{Spectral density at zero disorder}\label{app:zeroW}

In the absence of disorder, the cavity equations [Eq.~\eqref{eq:cavity}] become translationally invariant,  yielding a fixed cavity Green's function $G_0^{(\mu)}=\Delta$ on every node. Since the system is in the metallic phase at zero disorder, we can take the limit $\eta \to 0$. It is convenient to introduce the variable $X=\frac{\Delta}{1+t\Delta}$, in terms of which the self-consistent cavity equation reduces to
\begin{equation}
    X^{-1}=E+t-t^2k\frac{pX}{1-tpX} \, .
\end{equation}
This equation admits a complex solution within a finite energy interval $[\lambda_-, \lambda_+]$, corresponding to the continuous spectrum of the clean system. The band edges are given by
\begin{equation}\label{eq:E_lim}
    \lambda_\pm = t(p-1) \pm 2t\sqrt{kp} \, .
\end{equation}
Within this band, the retarded solution (with positive imaginary part) is,
\begin{equation}\label{eq:X}
    X=\frac{E + t + pt+ {\rm i}\sqrt{(E-\lambda_-)(\lambda_+-E)}}{2pt(E + t + kt)} \, .
\end{equation}
The on-site Green's function follows from the solution of $X$,
\begin{equation}\label{eq:G_self_W0}
    G(E)=\frac{1}{E-t^2(k+1)\frac{pX}{1-tpX}} \, .
\end{equation}
Substituting the expression for $X$ (Eq.~\eqref{eq:X}) and taking the imaginary part yields the density of states from $\rho (E)= \frac{1}{\pi} \Im G (E)$, 
\begin{equation}\label{eq:W0rho}
    \rho (E)= \frac{(k+1)\sqrt{(E-\lambda_-)(\lambda_+-E)}}{ 2 (E + t + k t) ((1 + k) p t-E)} \, .
\end{equation}

\begin{figure*}[t]
\centering
\includegraphics[width=0.75\linewidth]{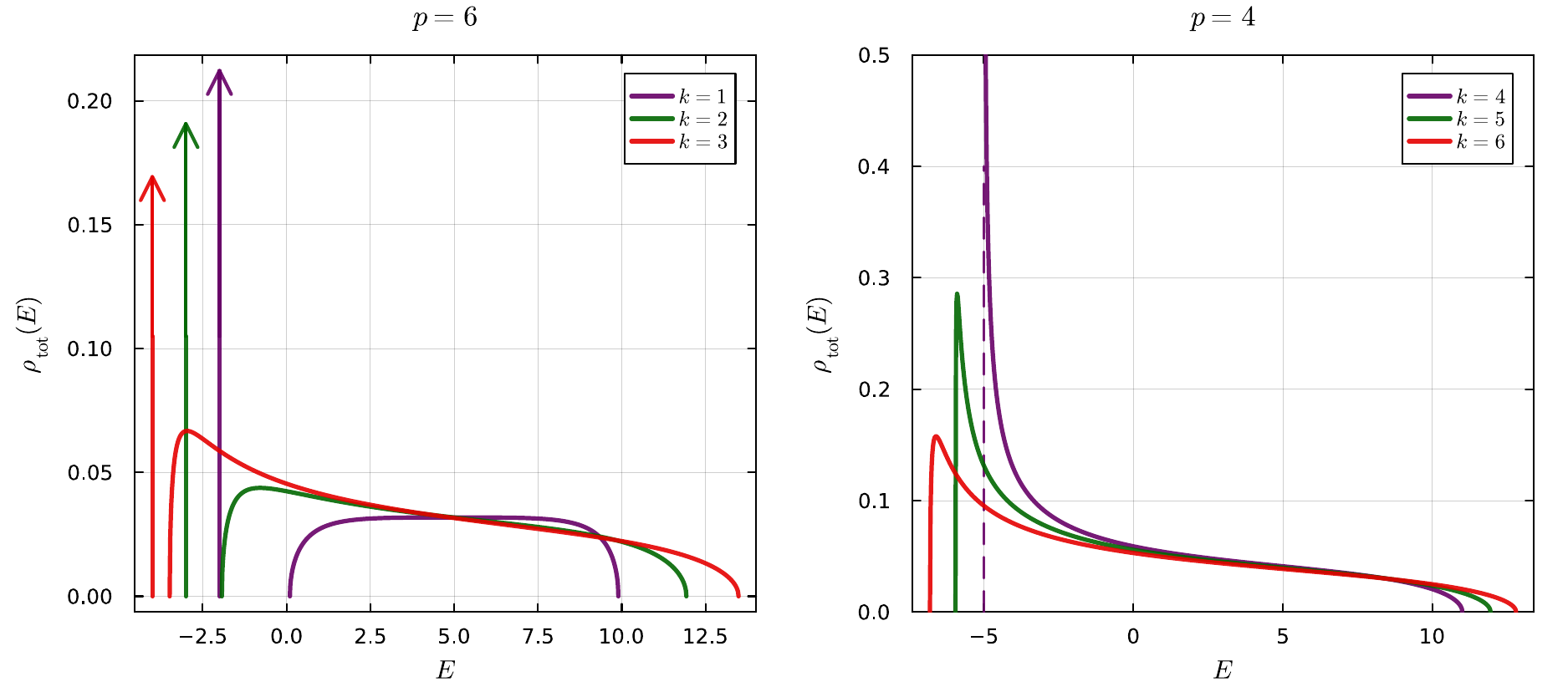}
\caption{Density of states at zero disorder ($W=0$) for $p=6$ (right) and $p=4$ (left) and several valuse of $k$. Solid lines show the continuous part of the spectrum for $k < p$, given by Eq.~\eqref{eq:W0rho}. Vertical arrows mark the positions of discrete eigenvalues at $E = -(k+1)t$; their height is proportional to the eigenvalue degeneracy density, $\frac{p-k}{p+1}$.}
\label{fig:MP_W0}
\end{figure*}

The imaginary part of Eq.~\eqref{eq:X} takes the form of a shifted and rescaled Marčenko–Pastur distribution~\cite{pastur1967distribution}. For large node-to-clique connectivity ($k\to\infty$) Eq.~\eqref{eq:W0rho} converges to the imaginary part of Eq.~\eqref{eq:X} and the  Marčenko–Pastur distribution is recovered. (For $p=1$, it reduces to the Wigner semicircle law~\cite{wigner1958distribution}, recovering the Bethe lattice result.) This spectral form emerges because at $W=0$, the Hamiltonian [Eq.~\eqref{eq:hamiltonian}] can be expressed as a sparse analog  of a Wishart matrix. To make this explicit, we introduce the incidence matrix of the clique-node structure, defined as a rectangular matrix $W_{i\mu}$ whose entries are equal to one whenever node $i$ belongs to clique $\mu$, and zero otherwise:
\begin{equation}
    W_{i \mu} = 
    \begin{cases}
        1 & \text{if } i \in \partial_\mu \Leftrightarrow \mu \in \partial_i \,,\\
        0 & \text{otherwise} \, .
    \end{cases}
\end{equation}
By construction, since each node belongs to $k+1$ cliques and each clique contains $p+1$ nodes, the row and column sums of $W$ satisfy
\begin{equation}
        \sum_{i=1}^N W_{i \mu} = p+1 \,, \qquad
    \sum_{\mu=1}^M W_{i \mu} = k+1 \, ,    
\end{equation}
with $\frac{M}{N} = \frac{k+1}{p+1}$. The matrix elements read
\begin{equation}
[W W^{T}]_{ij} = \sum_{\mu = 1}^M W_{i\mu} W_{j\mu}
= \sum_{\mu=1}^M \mathbf{1}_{i \in \partial_\mu}\,\mathbf{1}_{j \in \partial_\mu} \, .
\end{equation}
For $ i \neq j $ the $[W W^{T}]_{ij}$'s are equal to one if $ i $ and $ j $ belong to the same clique, and zero otherwise. The diagonal elements are equal to $ k+1 $, as they count the number of cliques to which a node belongs. The kinetic term of the tight-binding Hamiltonian on the Husimi tree can then be written simply as $H = t\, W W^{T} - t(k+1) \mathbb{I}$ ($\mathbb{I}$ being the $N \times N$ identity matrix), so that $H$ is a sparse random analog  of a Wishart matrix~\cite{kutlin2021emergent}.  The rectangularity parameter of the associated ensemble is $q = \frac{M}{N} = \frac{k+1}{p+1}$.

Hence, when $p>k$, the rank of $H$ exceeds the rank of $W$, and---as in the classical Wishart ensemble---Eq.~\eqref{eq:W0rho} describes only the bulk of the spectrum, satisfying $\int \rho(E)\, dE = q$ [the fact that $\rho(E)$ is not normalized to one is clearly visible from the plots of the DoS of Fig.~\ref{fig:MP_W0} for $p>k$], while the remaining fraction $N(1-q)$ of the eigenvalues form a degenerate set located at energy $-t(k+1)$ and separated from the bulk.  Hence, the total DoS at $W=0$ in the general case is
\begin{equation} \label{eq:rhoEtot}
\rho_{\rm tot} (E) = \left \{
\begin{array}{ll}
    \rho(E) + \frac{p-k}{p+1} \delta(E + t(k+1)) & \textrm{if~} p>k \, , \\
    \rho(E) & \textrm{if~} p \le k \, , \\
\end{array}
\right.
\end{equation}
where $\rho(E)$ is given in Eq.~\eqref{eq:W0rho}. Note that, as in the usual Marčenko–Pastur distribution for the Wishart ensemble, for $k=p$, i.e., $q=1$, the $\delta$-peak and the bulk part of the spectrum merge, which results in a square-root divergence of $\rho(E)$ at the lower edge (right panel of Fig.~\ref{fig:MP_W0}). Finally, there is also a single additional isolated eigenvalue at of energy $t(k+1)p$ corresponding to the uniform eigenvector $\psi = (1/\sqrt{N},\ldots,1/\sqrt{N})$~\cite{biroli2010anderson}.

The combination of the bulk and isolated spectral contributions for $p>k$ ensures that the average value of the energy's eigenstates is zero, since the Hamiltonian is traceless, i.e. $- (1-q)\, t (k+1) \;+\;\langle E \rangle_{\rho(E)} = 0$. The Marčenko–Pastur-like distribution of Eq.~\eqref{eq:W0rho}  thus satisfies
\begin{equation}
    \begin{aligned}
       \int \rho(E) dE &= \frac{k+1}{p+1} \, ,\\ 
       \int \rho(E) E dE &= \frac{t (p-k)(k+1)}{p+1} \, , 
       \end{aligned}
\end{equation}
if $p>k$, and 
\begin{equation}
    \begin{aligned}
       \int \rho(E) dE &= 1 \, , \qquad
       \int \rho(E) E dE = 0 \, ,
    \end{aligned}
\end{equation}
if $p \le k$. 

At $W=0$, the gap between the isolated part and the bulk for $p>k$ is $\lambda_- +t(k+1) = t (\sqrt{p} - \sqrt{k})^2$. When diagonal disorder is switched on, the $W=0$ degeneracy of the isolated states is lifted by disorder, producing a narrow band of states. The two continuous bands broaden, and the spectral edges are shifted by $\pm W/2$. Hence, upon increasing $W$, the two bands eventually merge at a disorder scaling as $t (\sqrt{p} - \sqrt{k})^2$, which is significantly smaller than the critical value at $E=0$ which scales as $W_c \simeq 4 kp \ln kp$.

\section{The cavity recursion equations}\label{app:cavity}

The Green’s function $G_{ij}(z)=\langle i | (z-\hat H)^{-1} | j \rangle$ admits a convenient formulation within a Gaussian field-theory framework.  
Introducing a vector field $\vec\phi$, the diagonal resolvent elements can be written as
\begin{equation}\label{eq:gaussianG}
    -{\rm i}\, G_{00}(z)
    = 
    \frac{\int [d\vec{\phi}] \, \phi_0^2 \, e^{S[\vec{\phi},z]}}
         {\int [d\vec{\phi}] \, e^{S[\vec{\phi},z]}} \, ,
\end{equation}
where the action 
\begin{equation} \label{eq:action}
 S[\vec{\phi},z]
= -\frac{{\rm i}}{2} \sum_{i,j=1}^N \big[z \delta_{ij} - \hat H_{ij}\big] \phi_i \phi_j   
\end{equation}
plays the role of an effective Gibbs-Boltzmann weight 
$P[\vec\phi]=e^{S[\vec{\phi},z]}/Z$, with $Z$ the normalization factor.
Hence, Eq.~\eqref{eq:gaussianG} identifies $-{\rm i}\,G_{00}(z)$ as the variance of the marginal distribution of $\phi_0$ after integrating out all other fields:
$P_0(\phi_0) \propto \exp\!\left[-\frac{{\rm i}}{2G_{00}} \phi_0^2 \right]$. We define the cavity Green’s function $G_0^{(\mu)}$ as the variance of $\phi_0$ when node $0$ is disconnected from the other nodes of one of the $(k+1)$ cliques to which it belongs, say clique $\mu$, $P_0^{(\mu)}(\phi_0) \propto 
\exp\!\left[-\frac{{\rm i}}{2 G_0^{(\mu)}} \phi_0^2 \right]$.
Because the action is quadratic, all cavity marginals remain Gaussian.

\medskip

To derive the self-consistent cavity equations, consider node $0$ with clique $\mu$ removed.  
Assume that all fields have been integrated out except those on nodes $i=1,\ldots,p$ belonging to the remaining $k$ cliques 
$\nu\in\partial_0\setminus\mu$. We assume that the marginal probability distributions on these variables is completely factorized, due to the fact that these nodes belong to different branches of the tree that becomes completely disconnected in the thermodynamic limit. The resulting measure therefore contains only the marginal distributions of these fields and the local action terms coupling them to $\phi_0$:
\begin{widetext}
\begin{equation}\label{eq:P0mu}
\begin{aligned}
P_0^{(\mu)}[\phi_0]
&\propto 
e^{-\tfrac{{\rm i}(z-\epsilon_0)}{2}\phi_0^2}
\prod_{\nu\in\partial_0\setminus\mu}
\Bigg[
\int \prod_{i\in\partial_\nu\setminus 0} d\phi_{i,\nu}\;
\exp\Bigg(
{\rm i} t \phi_0 \sum_{i=1}^p \phi_{i,\nu}
+\frac{{\rm i} t}{2} \sum_{i\neq j=1}^p \phi_{i,\nu}\phi_{j,\nu}
-\frac{{\rm i}}{2} \sum_{i=1}^p \frac{\phi_{i,\nu}^2}{G_i^{(\nu)}}
\Bigg)
\Bigg] \, .
\end{aligned}
\end{equation}
One may evaluate these Gaussian integrals either by explicitly inverting the $p\times p$ interaction matrix---whose off-diagonal elements are all equal to $t$ and whose diagonal elements are equal to $1/G_i^{(\nu)}$---which describes the Gaussian weight over the variables $\phi_{i,\nu}$'s (this approach is presented in Appendix ~\ref{app:cavity_matrix} below), or by introducing an auxiliary variable 
$m_\nu=\sum_{i=1}^p\phi_{i,\nu}$
via the identity
$\prod_{\nu\in\partial_0\setminus\mu} \int d m_\nu\, \delta\!\left(m_\nu-\sum_{i=1}^p\phi_{i,\nu}\right)=1$.
Using the Fourier representation of the $\delta$ functions, Eq.~\eqref{eq:P0mu} becomes
\begin{equation} \label{eq:P0mu2}
\begin{aligned}
            P_0^{(\mu)}[\phi_0] & \propto e^{-\tfrac{{\rm i}(z-\epsilon_0)}{2}\phi_0^2}  \prod_{\nu\in\partial_0\setminus\mu} \Bigg \{ \int \frac{d m_\nu \, d \hat{m}_\nu}{2 \pi} \times \\
            & \qquad \qquad \; \; \times 
            \prod_{i\in\partial_\nu \setminus 0} d \phi_{i,\nu} \exp \left[ {\rm i} \hat{m}_\nu \left( m_\mu - \sum_{i=1}^p \phi_{i,\nu} \right) + {\rm i} t \phi_0 m_{\nu} + \frac{{\rm i} t}{2} m_{\nu}^2 - \frac{\rm i}{2} \sum_{i=1}^p \left( \frac{1}{G_i^{(\nu)}} + t \right) \phi_{i,\nu}^2 \right] \Bigg \} \\
            & \propto \exp \left [ -\tfrac{{\rm i}}{2} \left( z-\epsilon_0 - t^2 \sum_{\nu\in\partial_0\setminus\mu} \frac{\sum_{i\in\partial_\nu \setminus 0} \frac{G_i^{(\nu)}}{1 + t G_i^{(\nu)} }  } {1 - t \sum_{i\in\partial_\nu \setminus 0} \frac{G_i^{(\nu)}}{1 + t G_i^{(\nu)} } }  \right) \phi_0^2 \right] \, .
            \end{aligned}
\end{equation}
The last equality is obtained by successively performing the Gaussian integrals over $\phi_{i,\nu}$, $\hat m_\nu$, and $m_\nu$, and gives the cavity recursion relation, Eq.~\eqref{eq:cavity} (with $z = {\rm i}\eta$ since we specifically focus on the case $E = 0$).  The expression for the full Green’s function $G_{00}$, Eq.~\eqref{eq:G}, follows from the same procedure, when computing the marginal probability distribution in presence of all the neighbors. 
\subsection{Derivation using matrix inversion}\label{app:cavity_matrix}
An alternative derivation of these equations is which does not rely on the introduction of the auxiliary collective clique variables (and their Fourier conjugates) $m_\nu$ and $\hat m_\nu$ is possible. The starting point is Eq.\eqref{eq:P0mu}, which gives the local measure with marginal cavity Gaussian weights for the fields $\phi_{i,\nu}$ associated with the nodes belonging to the remaining $k$ cliques $\nu \in \partial_0 \setminus \mu$ of a given node $0$, after removing one clique $\mu$. This marginal Gaussian weight can be written compactly in terms of $p \times p$ matrices $A^{(\nu)}$ and vectors $B^{(\nu)}$ as
\begin{equation} \label{eq:PAB}
\begin{aligned}
    P_0^{(\mu)}[\phi_0] &\propto e^{-\tfrac{{\rm i}(z-\epsilon_0)}{2}\phi_0^2} \prod_{\nu\in\partial_0\setminus\mu} \Bigg[ \int \prod_{i\in\partial_\nu\setminus 0} d\phi_{i,\nu}\; \exp\Bigg(
-\frac{{\rm i}}{2} \sum_{i,j=1}^p \phi_{i,\nu} A_{ij}^{(\nu)} \phi_{i,\nu} + {\rm i} \sum_{i=1}^p B_i^{(\nu)} \phi_{j,\nu}
\Bigg)
\Bigg] \\
& = \exp \left[ -\tfrac{{\rm i}}{2} \left( z-\epsilon_0 - t^2 \sum_{\nu\in\partial_0\setminus\mu} \sum_{i,j=1}^p \left[ A^{(\nu)} \right]^{-1}_{ij} \right) \phi_0^2 \right]
\, .
\end{aligned}
\end{equation}
\end{widetext}

The matrices $A^{(\nu)}$ and vectors $B^{(\nu)}$ are defined as
\begin{equation} \label{eq:AB}
\begin{aligned}
A_{ij}^{(\nu)}
&= \frac{1}{G_i^{(\nu)}} \delta_{ij} - t (1-\delta_{ij}) =
\begin{cases}
\displaystyle \frac{1}{G_i^{(\nu)}} & \text{if } i=j \, , \\[1mm]
-t & \text{if } i \neq j \, ,
\end{cases}
\\[2mm]
B_i^{(\nu)} &= t \phi_0 \, .
\end{aligned}
\end{equation}
Therefore, to obtain the cavity recursion relation, it suffices to invert the matrices $A^{(\nu)}$. Remarkably, this inversion can be performed exactly for the specific structure in Eq.~\eqref{eq:AB}. To simplify notation, we drop the superscript $(\nu)$ in what follows.

We consider a generic $p \times p$ matrix $A$ of the form above and denote its inverse by $M$, satisfying $\sum_{j=1}^p M_{ij} A_{jk} = \delta_{ik}$. This yields 
\begin{equation} \label{eq:inverse}
\frac{M_{ik}}{G_{i}} - t \sum_{j \neq k} M_{ij} = \delta_{ik} \, .
\end{equation}
Introducing $S_i = \sum_{j \neq i} M_{ij}$ and using $\sum_{j \neq k} M_{ij} = S_i + M_{ii} - M_{ik}$, Eq.~\eqref{eq:inverse} leads to the coupled equations for the diagonal and off-diagonal elements of $M$:
\begin{equation} \label{eq:M1}
\left \{
\begin{array}{ll}
\dfrac{M_{ii}}{G_i} - t S_i = 1 \, , \\[2mm]
M_{ik} \! \left(\dfrac{1}{G_k}+t\right) - t(S_i+M_{ii}) = 0 \, ,
\qquad i \neq k \, .
\end{array}
\right.
\end{equation}
Defining $\chi_i^{-1} = 1/G_i + t$, as in the main text, the second equation becomes
\begin{equation} \label{eq:M2}
    \frac{M_{ik}}{\chi_k} - \frac{M_{ii}}{\chi_i} + 1 = 0 \, , \qquad \textrm{for~} i \neq k \, .
\end{equation}
Summing over all $k$ different from $i$, we obtain
\begin{equation}
    S_i = \sum_{k \neq i} M_{ik} = \left( \frac{M_{ii}}{\chi_i} - 1 \right) \sum_{k \neq i} \chi_k \, .
\end{equation}
From the first line of Eq.~\eqref{eq:M1}, one has $M_{ii}/\chi_i = t ( S_i + M_{ii}) + 1$. Substituting this into the equation above yields $S_i = t(S_i + M_{ii}) \sum_{k \neq i} \chi_k$, which can be solved to give
\begin{equation}
    S_i = \frac{t M_{ii} \sum_{k \neq i} \chi_k }{1 - t \sum_{k \neq i} \chi_k} \, .
\end{equation}
Inserting this result back into Eq.~\eqref{eq:M1}, after straightforward algebra one obtains
\begin{equation}
\begin{aligned}
    M_{ii} &= \frac{G_{i} \left( 1 - t \sum_{k \neq i} \chi_k \right)}{1 - t \left[ \left( 1 + t G_{i} \right) \sum_{k} \chi_k  - G_{i} \right] } \, , \\
    S_i &= \frac{t G_{i} \sum_{k \neq i} \chi_k}{1 - t \left[  \left( 1 + t G_{i} \right)\sum_{k} \chi_k - G_{i} \right] } \, .
\end{aligned}
\end{equation}
The contribution of a neighboring clique to the local cavity Green’s function at node $0$ follows directly from Eq.~\eqref{eq:PAB} and is proportional to 
\begin{equation}
\begin{aligned}
    \sum_{i,j}^p M_{ij} &= \sum_{i=1}^p \left( M_{ii} + S_i \right)\\ 
    &= \sum_i \frac{G_{i}}{1 - t \left[  \left( 1 + t G_{i} \right)\sum_{k} \chi_k - G_{i} \right] } \\
    &= \sum_{i=1}^p \frac{G_{i}}{\left( 1 + t G_{i} \right) \left( 1 - t \sum_k \chi_k\right)}\\
    &= \sum_{i=1}^p \frac{\chi_i} {1 - t \sum_k \chi_k} \, .
\end{aligned}
\end{equation}
Finally, inserting this identity into Eq.~\eqref{eq:PAB}, we obtain the self-consistent recursion relation for the cavity Green’s functions,
\begin{equation}
    \left [ G_{0}^{(\mu)} \right] ^{-1} = z - \epsilon_0 - t^2 \sum_{\nu\in\partial_0\setminus\mu} \frac{\sum_{i \in \partial_\nu \setminus 0} \chi_i^{(\nu)}}{1 - t \sum_{i \in \partial_\nu \setminus 0} \chi_i^{(\nu)} } \, ,
\end{equation}
which, upon expressing $\chi_i^{(\nu)} = G_i^{(\nu)}/(1+tG_i^{(\nu)})$, exactly coincides with Eq.~\eqref{eq:cavity}.

\section{The linear stability analysis}\label{app:operator}

As discussed in the main text, Anderson localization follows from the linear stability of Eq.~\eqref{eq:cavity} with respect to a small imaginary component. Introducing the variable $\chi_i^{(\mu)}=\frac{G_i^{(\mu)}}{1+tG_i^{(\mu)}}$, setting $\chi_{i}^{\mu}=x_{i}^{\mu}+i\eta\,\hat x_{i}^{\mu}$, and expanding Eq.~\eqref{eq:cavity} to first order in $\eta$ yields the linear recursion relations~\eqref{eq:Xreal_self}–\eqref{eq:Ximaginary_self}. After averaging over disorder realizations, these equations define a self-consistent integral equation for the joint probability density of $(x_i,\hat{x}_i)$. For simplicity, we now focus on $k=1$ (the general case is considered in Appendix \ref{app:operator_general} below) and set $t=1$. Due to the linearity of the recursion in $\hat{x}_i$, the self-consistent equation for the partial Fourier transform with respect to the second argument reads~\cite{abou1973selfconsistent}
\begin{equation} \label{eq:Pxxh}
\begin{aligned}
    \hat{P} (x, \lambda) & = \int d \gamma(\epsilon) \prod_{i=1}^p \hat{P} \left ( x_i, \frac{\lambda x^2}{(1 - \sum_{i=1}^p x_i)^2} \right) \times \\
    & \qquad \;\; \times \delta \left( x^{-1} -  1 + \epsilon_0 + \frac{\sum_{i=1}^p x_i}{1- \sum_{i=1}^p x_i} \right) \, .
    \end{aligned}
\end{equation}
Assuming that for large $\hat{x}$ (i.e. small $\lambda$) the distribution of imaginary parts develops power-law tails of the form given in Eq.~\eqref{eq:ansatz}~\cite{abou1973selfconsistent,mirlin1991localization,tikhonov2019critical}, $\hat{P}(x,\lambda)\simeq P_0(x)-f(x)|\lambda|^\beta$, with $P_0(x)$ the distribution of real parts satisfying Eq.~\eqref{eq:Xreal_self}, a small-$\lambda$ expansion leads to a self-consistency equation identifying $f(x)$ as the eigenfunction of the linear operator~\eqref{eq:operator}, with kernel
\begin{equation}
    \label{eq:kernel}
    \begin{aligned}
    K_\beta(\tilde x, x) & =\frac{p}{|x|^{2 - 2 \beta}} \int d \gamma(\epsilon) \, R \left( \frac{x^{-1} - \epsilon - 1}{x^{-1} - \epsilon - 2} - \tilde{x} \right) \times \\
    & \qquad \qquad \qquad \times \left \vert x^{-1} - \epsilon - 2 \right \vert^{2 \beta-2}  \, . 
    \end{aligned}
\end{equation}
Here $R(y)$ is the distribution of the sum of $p-1$ real parts $x_i$ [see Eq.~\eqref{eq:RS} below], which is Cauchy for large $p$, $R(y) \simeq \frac{a}{\pi ( y^2 + a^2)}$, 
with $a \propto (p-1)/W$. Since $W_c\simeq 4p\ln p$ [see Eq.~\eqref{eq:Wc}], near criticality one has $a\sim 1/\ln p$. 
From the structure of the operator, Eqs.~\eqref{eq:operator} and~\eqref{eq:kernel}, the eigenfunction $f_\beta(x)$  takes the form [see Fig.~\ref{fig:lambda}(b)]
\begin{equation} \label{eq:f}
     f_\beta(x) \simeq 
    \left \{
    \begin{array}{ll}
     c_\beta |x|^{2 \beta-2} & \textrm{for~} |x| \gg 2/W \, , \\
    f_\beta (0) & \textrm{for~} |x| \ll 2/W \, . \\
    \end{array}
    \right.
\end{equation}
Inserting the ansatz~\eqref{eq:f} into the operator in the regimes $|x|\gg 2/W$ and $|x|\ll 2/W$ yields the coupled equations
\begin{widetext}
    \begin{align} 
    c_\beta &= \frac{8 p}{W} \, \lambda_\beta(W) \int_{0}^{+\infty} \frac{d s}{2 \pi} \cos(s) \, e^{-a s} \,  F_{\beta} \left( s, \frac{2}{W} \right) \left[ f_\beta (0) \,  \frac{\sin \left( \frac{2 s}{W} \right) }{s} + c_\beta F_{1 - \beta} \left( s, \frac{2}{W} \right) \right] \,\label{eq:c} ,\\
    f_\beta (0) &= 4 p \, \lambda_\beta(W) \int_{0}^{+\infty} \frac{d s}{2 \pi} \cos(s) \, e^{-a s} \left[ f_\beta (0) \,  \frac{\sin \left( \frac{2 s}{W} \right) }{s} + c_\beta F_{1 - \beta} \left( s, \frac{2}{W} \right) \right] \, ,\label{eq:f0}
\end{align}
where the function $F_\beta$ is defined as
\begin{align} \label{eq:Fbeta}
    F_{\beta} (s,\sigma) &= \int_{\sigma}^{+\infty} d y \frac{\cos(sy)}{y^{2 \beta}}
    \simeq s^{2 \beta - 1} \Gamma(1 - 2 \beta) \sin (\pi \beta) + \frac{\sigma^{1-2 \beta}}{2 \beta - 1} \, .
\end{align}
Using the expansion~\eqref{eq:Fbeta} and neglecting $O(a)$ corrections (subleading as $a\sim 1/\ln p$ for $p$ large), one recovers Eq.~\eqref{eq:lambda} for $k=1$. 

\subsection{General case of the linear integral operator for arbitrary $k$} \label{app:operator_general}

In this section we compute the largest eigenvalue of the linear integral operator governing the stability of the cavity recursion relations with respect to a small imaginary perturbation, for the general case $k>1$.  
Our starting point is the set of linearized recursion relations given in Eqs.~\eqref{eq:Xreal_self} and \eqref{eq:Ximaginary_self}.  
The self-consistent equation for the joint probability distribution of the variables $x_i^{\mu}$ and $\hat x_i^{\mu}$, partially Fourier transformed with respect to the second argument---see Eq.\eqref{eq:Pxxh}---generalized to an arbitrary value of $k$, reads:
\begin{equation} \label{eq:Pxl_kp}
\begin{aligned}
    \hat{P} (x, \lambda) & = \int d \gamma(\epsilon)  \prod_{\nu=1}^k \left [ \prod_{i=1}^p dx^{(\nu)}_i\hat{P} \left ( x^{(\nu)}_i, \frac{\lambda t^2x^2}{1 - t\sum_{l=1}^px^{(\nu)}_l} \right) \right] \delta \left( x^{-1} -  t + \epsilon_0 +  t^2\sum_{\nu=1}^k\frac{\sum_{i=1}^p x^{(\nu)}_i}{1- t\sum_{i=1}^px^{(\nu)}_i} \right) \, .
    \end{aligned}
\end{equation}
We assume that at large $\hat{x}$ (corresponding to small $\lambda$) the distribution of imaginary parts develops a power-law tail~\cite{abou1973selfconsistent,mirlin1991localization,tikhonov2019critical}, $\hat{P}(x,\lambda)\simeq P_0(x)-f(x)|\lambda|^\beta$, where $P_0(x)$ is the distribution of the real parts, which can be shown to follow a Cauchy distribution for large $p$. Inserting this ansatz into Eq.~\eqref{eq:Pxl_kp} and expanding it to leading order in $\lambda$ yields an eigenvalue equation for $f(x)$:
\begin{equation}
    \label{eq:operator_kp}
    f_\beta(x) = \frac{p k }{|x/t|^{2 - 2 \beta}} \int d \gamma(\epsilon) \, d S(w) \,  R \left( \frac{x^{-1} - \epsilon + t^2 w - t}{t(x^{-1} - \epsilon + t^2 w - 2t)} - \tilde{x} \right) \left \vert x^{-1} - \epsilon + t^2 w - 2t \right \vert^{2 \beta-2} f_\beta (\tilde{x}) \, d \tilde{x} \,.
\end{equation}
\end{widetext}
Here, $R(y)$ denotes the distribution of the sum of the $p-1$ real parts $x_i^{(\mu)}$, which follows a Cauchy distribution, while the auxiliary distribution $S(w)$ is defined as the distribution of the sum of $k-1$ independent random variables, each expressed as a function of $x_i^{(\mu)}$:
\begin{equation} \label{eq:RS}
\begin{aligned}
        R(y) & = \int \prod_{i=2}^p d P_0 (x_{i}^{(\mu)} ) \, \delta \left ( y - \sum_{i=2}^p x_{i}^{(\mu)} \right) \, , \\
        S(w) & = \int \prod_{\mu=2}^k d P_0 (x_{i}^{(\mu)} ) \, \delta \left ( w - \sum_{\mu=2}^k \frac{\sum_{i=1}^p x_{i}^{(\mu)}}{1 - t \sum_{i=1}^p x_{i}^{(\mu)}} \right) \, .
        \end{aligned}
\end{equation}
In the large-$k$ limit one can show that the distribution $S(w)$ also becomes Cauchy. The typical value of the sum of $p-1$ variables $x_i^{(\mu)}$ is simply $(p-1)$ times the typical value of $x_i^{(\mu)}$, and the typical value of $w$ is $(k-1)$ times the same typical value. One thus has that asymptotically, for large $k$ and $p$:
\begin{equation}
    \begin{aligned}
        R(y) & \simeq \frac{a}{\pi ( y^2 + a^2)} \, , \qquad
        S(w) & \simeq \frac{b}{\pi ( w^2 + b^2)} \, .
    \end{aligned}
\end{equation}
Since the typical value of $x_i^{(\mu)}$ scales as $1/W$, it follows immediately that  $a \propto (p-1)/W$ and $b \propto (k-1)/W$. Moreover, as shown below [and stated in Eq.~\eqref{eq:Wc}], the critical disorder scales as $W_c \propto 4kp \ln(kp)$. Thus close to the critical point both $a$ and $b$ vanish respectively as $(k \ln(kp))^{-1}$ and $(p \ln(kp))^{-1}$, and therefore give only subleading contributions to the leading term of the largest eigenvalue.

To see this explicitly, we split the integral over $w$ in Eq.~\eqref{eq:operator_kp} into two regions,  $|w| < C b$ and $|w| > C b$, where $C$ is a constant of order one (with $C>1$).  In the first region, $|w| < C b$, the $w$ dependence in the argument of $R$ as well as in the absolute in value Eq.~\eqref{eq:operator_kp} can be neglected. We can then integrate over $dw$ and using  $\int_{|w|< C b} S(w) \, dw \approx 1$ for $C$ large enough (since $b \ll C$), the integral finally reproduces the same expression of the Kernel obtained in the $k=1$ case [see Eq.~\eqref{eq:kernel}], the only difference being that the prefactor $p$ is replaced by $kp$. In the second region, $|w| > C b$, the value of $w$ cannot be neglected; however, for sufficiently large $C$ the distribution satisfies $S(w) \ll 1$ in this interval. Consequently, this contribution is subleading and gives only a contribution of $O((p \ln(kp))^{-1})$ close to criticality and can be safely neglected at leading order for large $p$.

Therefore, upon making the simple replacement of the prefactor in front of the integral, $p \to kp$, the above analysis---together with the detailed $k=1$ computation above---directly yields the expression for the largest eigenvalue of the integral operator reported in Eq.~\eqref{eq:lambda}.

\section{Correlation function}\label{app:corr}

\begin{figure}
    \centering
    \includegraphics[width=0.95\linewidth]{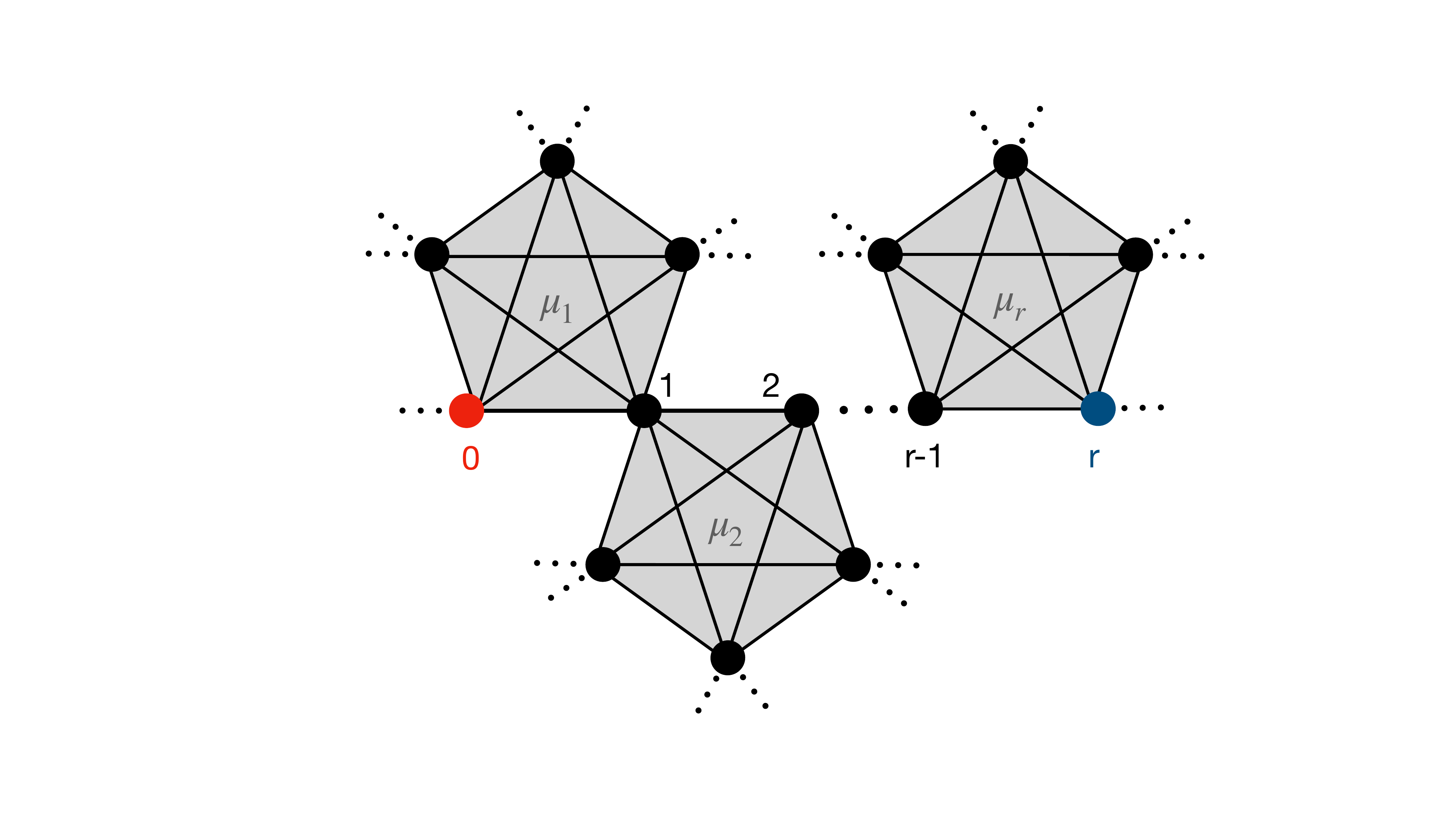}
    \vspace{-0.6cm}
    \caption{
        Subgraph relevant to paths connecting nodes $0$ and $r$ in the Husimi tree for $k=1$ and $p=4$.  
        The shortest-path nodes are labeled $i=0,\dots,r$, and $\mu_i$ denotes the clique shared by nodes $i-1$ and $i$.
    }
    
    \label{fig:corr_path}
\end{figure}

We derive the off-diagonal Green’s function $G_{0r}(z)$ between two nodes $0$ and $r$ separated by $r$ cliques along the Husimi tree. Because of the underlying tree structure, the shortest path between them is unique and traverses the sequence of nodes $i=0,1,\dots,r$.  Each consecutive pair $(i-1,i)$ belongs to a single clique, denoted $\mu_i$. Neighboring cliques share exactly one node, which therefore necessarily belongs to any path connecting $0$ and $r$. As a consequence, the nodes $i=0,1,\dots,r$ are common to all paths between the two endpoints. Unlike on the Bethe lattice, however, the internal structure of each clique generates multiple distinct paths of finite length that contribute to $G_{0r}(z)$ once the endpoints are fixed. Figure~\ref{fig:corr_path} illustrates the subgraph that contributes to the paths between nodes $0$ and $r$ for the case $k = 1$ and $p = 4$. The correlation function $C(r)$ is given by the average of the squared off-diagonal Green’s function, $C(r) = \langle |G_{0r}|^2 \rangle$ and is related to the correlation of the eigenfunctions' amplitudes as 
\begin{equation}
\left \langle |\psi_\alpha(i)|^2 |\psi_\alpha(i+r)|^2  \delta(E_\alpha) \right \rangle = \lim_{\eta \to 0^+} \frac{\eta \, C(r)} {\langle \mathrm{Im}\,G_{ii} \rangle} \, ,
\end{equation}
where the average is performed over all pairs of nodes at distance $r$, all eigenstates of zero energy, and all disorder realizations.

Using the Gaussian field representation already introduced in Appendix ~\ref{app:cavity}, we express the two-point Green's function as
\begin{equation}\label{eq:G0r_gaussian}
    -{\rm i}\, G_{0r}(z)=\frac{\int [d\vec{\phi}]\, \phi_0 \phi_r \, e^{S[\vec{\phi},z]}} {\int [d\vec{\phi}]\, e^{S[\vec{\phi},z]}} \, ,
\end{equation}
where the action $S[\vec{\phi},z]$ is given in Eq.~\eqref{eq:action}.
The numerator is evaluated by successively integrating the fields
associated with the $r$ cliques
$\mu_1,\mu_2,\ldots,\mu_r$ along the unique shortest path.
As in the diagonal case, all contributions from fields outside this path cancel identically between numerator and denominator.

We begin with the integration over the first clique $\mu_1$.
In the numerator, all fields belonging to $\mu_1$ except $\phi_0$ and $\phi_1$ appear only quadratically, and their Gaussian integrals cancel against the denominator.
To decouple the internal clique interactions, we introduce the auxiliary variable $m_{\mu_1}=\sum_{i=2}^p \phi_{i,\mu_1}$,
representing the collective mode of clique $\mu_1$ once the boundary nodes $0$ and $1$ are excluded, together with its Fourier conjugate $\hat m_{\mu_1}$.
The resulting contribution to the numerator coming from the variables belonging clique $\mu_1$ except the node $1$ which is along the path reads
\begin{widetext}
\begin{align} \label{eq:I_mu1_raw} 
    \mathcal{I}_{\mu_1}^{(1)} &= \int d\phi_0\, \phi_0 \,  dm_{\mu_1}\, d\hat m_{\mu_1} \prod_{i=2}^p d\phi_{i,\mu_1} \, \exp \! \Bigg[ {\rm i}\hat m_{\mu_1}\! \left(m_{\mu_1}-\sum_{i=2}^p \phi_{i,\mu_1}\right) +{\rm i} t \phi_0 ( \phi_1 + m_{\mu_1} ) +\frac{{\rm i} t}{2} (\phi_1 + m_{\mu_1})^2 \nonumber \\
    &\qquad\qquad\qquad -\frac{{\rm i}}{2} \sum_{i=2}^p\! \left(\frac{1}{G_i^{(\mu_1)}}+t\right)\phi_{i,\mu_1}^2 -\frac{{\rm i}}{2}\frac{\phi_0^2}{G_{0}^{(\mu_1)}} -\frac{{\rm it}}{2} \phi_{1,\mu_1}^2  -\frac{{\rm i}}{2}(z-\epsilon_1)\phi_1^2 \Bigg] \, .
\end{align}
Performing first the Gaussian integrals over $\phi_{i,\mu_1}$ with $i\ge2$ and over $\hat m_{\mu_1}$ removes these variables and produces the same factors as in Eq.~\eqref{eq:Pxxh}, but now with a double cavity inside the clique. As done above and in the main text, it is convenient to introduce $\chi_0^{(\mu_1)}=G_0^{(\mu_1)}/(1+t G_0^{(\mu_1)})$. We also define 
\begin{equation}
    \Gamma_{\mu_1}^{(0,1)} =\frac{\sum_{i\in\partial_{\mu_1}\setminus\{0,1\}} \chi_i^{(\mu_1)}}{1-t\sum_{i\in\partial_{\mu_1}\setminus\{0,1\}} \chi_i^{(\mu_1)}} 
\end{equation} 
as the variance associated with the auxiliary variable $m_{\mu_1}$ after the integration over the $\phi_{i,\mu_1}$'s. Integrating over $m_{\mu_1}$ yields
\begin{align}
    \label{eq:I_mu1_after_m}
    \mathcal{I}_{\mu_1}^{(1)}= \int d\phi_0\, \phi_0 \exp\!\Bigg[ - \frac{{\rm i}}{2} \biggl(\frac{1}{G_{0}^{(\mu_1)}} - t^2 \Gamma_{\mu_1}^{(0,1)}\biggr) \phi_0^2 -\frac{{\rm i}}{2}(z-\epsilon_1-t^2 \Gamma_{\mu_1}^{(0,1)})\,\phi_1^2 + {\rm i}t(1+t\Gamma_{\mu_1}^{(0,1)})\phi_0\phi_1 \Bigg] \, .
\end{align}
From the definitions of the variances it follows that 
\begin{equation}\label{eq:Gamma_double}
 \frac{1}{\Gamma_{\mu}^{(a,b)}} = \frac{1}{\sum_{i\in\partial_{\mu}\setminus\{a,b\}}  \chi_i^{(\mu)}} - t = \frac{1}{\sum_{i\in\partial_{\mu}\setminus a }  \chi_i^{(\mu)} -  \chi_b^{(\mu)} } - t = \frac{1}{\frac{\Gamma_{\mu}^{(a)}}{1 + t \Gamma_{\mu}^{(a)}} - \frac{G_{\mu}^{(b)}}{1 + t G_{\mu}^{(b)} } } - t \, ,
 \end{equation}
\end{widetext}
which gives the identity
\begin{equation}
    \Gamma_{\mu}^{(a,b)}= \frac{\Gamma_{\mu}^{(a)} - G_b^{(\mu)}}
     {1 + t G_b^{(\mu)}(2 + t\Gamma_\mu^{(a)})} \, .
\end{equation}
Integrating over $\phi_0$ and using this identity, after some simple algebraic manipulations the result can be rewritten entirely in terms of single-cavity quantities. The remaining Gaussian integral over $\phi_0$ gives,
\begin{equation}\label{eq:I_mu1_final}
    \mathcal{I}_{\mu_1}^{(1)} = t\, \chi_0^{(\mu_1)}\,\bigl(1+t\Gamma_{\mu_1}^{(1)}\bigr)\, \phi_1\,\exp\!\left[ -\frac{{\rm i}}{2} (z-\epsilon_1 - t^2 \, \Gamma_{\mu_1}^{(1)})\,\phi_1^2 \right] \, .
\end{equation}
Equation~\eqref{eq:I_mu1_final} shows that integrating out clique $\mu_1$ transfers the correlation from $\phi_0$ to $\phi_1$, producing a multiplicative factor $\phi_1t\,\chi_0^{(\mu_1)}(1+t\Gamma_{\mu_1}^{(1)})$. Integrating over all the $\phi_i$'s associated with the other $k-1$ cliques $\mu_a$ to which node $1$ belongs---excluding the nodes of the clique $\mu_2$, which lies along the path---produces an additional contribution to the inverse variance of $\phi_1$, equal to $\Gamma_{\mu_a}^{(1)}$. Summing all these contributions, and using the self-consistent recursion equations for the cavity Green's functions, Eq.~\eqref{eq:cavity}, the inverse variance of $\phi_1$ is given by
\begin{equation}
z - \epsilon_1 - t^2 \, \Gamma_{\mu_1}^{(1)} 
- t^2 \!\! \sum_{\mu_a \in \partial_1 \setminus \{\mu_1,\mu_2\}} \! \! \Gamma_{\mu_a}^{(1)}
= \left( G_1^{(\mu_2)}\right)^{-1} \, .
\end{equation}
Therefore, the integration over the next clique $\mu_2$ proceeds identically, with $\phi_1$ and $\phi_2$ replacing $\phi_0$ and $\phi_1$. Iterating this procedure along the chain of cliques $\mu_1,\ldots,\mu_r$ yields the product
\begin{equation}\label{eq:prod_path}
    \prod_{\ell=1}^r t\,\chi_{\ell}^{(\mu_{\ell+1})} \bigl(1+t\Gamma_{\mu_{\ell+1}}^{(\ell+1)}\bigr) \, .
\end{equation}
The integration over the last clique $\mu_r$ produces a multiplicative factor $t\,\chi_{r-1}^{(\mu_r)}(1+t\Gamma_{\mu_r}^{(r)})$ and yields a linear factor $\phi_r$. This factor combines with the original $\phi_r$ already present in the numerator of Eq.~\eqref{eq:G0r_gaussian}, so that the final expression contains $\phi_r^2$. At this stage, node $r$ has no cavities in its neighbors, and its Gaussian weight is therefore governed by the full Green’s function $G_{rr}$. The remaining integral is
\begin{equation}
    \int d\phi_r\, \phi_r^2 \exp\!\left[-\frac{{\rm i}}{2G_{rr}}\,\phi_r^2\right] = -{\rm i}\, G_{rr} \, ,
\end{equation}
which coincides with the diagonal resolvent at site $r$. Combining this result with Eq.~\eqref{eq:prod_path} we finally obtain
\begin{equation}\label{eq:G0r_final}
    G_{0r} =  G_{rr}\prod_{\ell=1}^r \left[ t \chi_{\ell-1}^{(\mu_\ell)} \left(1+t\Gamma_{\mu_\ell}^{(\ell)}\right) \right] \, .
\end{equation}
The expression for the off-diagonal Green's function, Eq.~\eqref{eq:G0r_final}, factorizes into a product of (strongly correlated) random variables defined in terms of the local Green's functions along the unique clique path. Each factor, 
\begin{equation}
T_{\ell-1}^{(\mu_\ell)} \equiv t \chi_{\ell-1}^{(\mu_\ell)}(1+t\Gamma_{\mu_\ell}^{(\ell)}) =  \frac{t \chi_{\ell-1}^{(\mu_\ell)}}{1 - t \sum_{i \in \mu_\ell \setminus \ell} \chi_i^{(\mu_\ell)}} \, ,
\end{equation}
is strongly correlated with the successive term along the path via the cavity recursion relations. The correlation function is finally written as
\begin{equation} \label{eq:C0r}
    C_{0r} =  \left \langle \left |G_{rr} \right |^2 \prod_{\ell=1}^r \left | T_{\ell-1}^{(\mu_\ell)} \right|^2 \right \rangle \, .
\end{equation}

We now specialize to the localized phase, in which the Green’s functions---and therefore the variables $\chi_i^{(\mu)}$---can be linearized with respect to a small imaginary part as $\chi_i^{(\mu)} = x_i^{(\mu)} + i \eta\, \hat{x}_i^{(\mu)}$ where $x_i^{(\mu)}$ and $\hat{x}_i^{(\mu)}$ satisfy Eqs.~\eqref{eq:Xreal_self} and \eqref{eq:Ximaginary_self}. In the limit
$\eta \to 0$, the factors $T_{\ell-1}^{(\mu_\ell)}$ become
\begin{equation}
T_{\ell-1}^{(\mu_\ell)} =
\frac{t\, x_{\ell-1}^{(\mu_\ell)}}{1 - t \sum_{i \in \mu_\ell \setminus \ell} x_i^{(\mu_\ell)}} + O(\eta) \, .
\end{equation}
By comparing Eq.~\eqref{eq:C0r} with Eq.~\eqref{eq:Ximaginary_self}, one immediately sees that---similarly to the Anderson model on the infinite loopless Bethe lattice---the square of the real part of $T_{\ell-1}^{(\mu_\ell)}$ given above coincides with the multiplicative factors appearing in front
of the $\hat{x}_\ell^{(\mu)}$'s. As a consequence, in the linearized regime, the equation for the imaginary part of the Green’s function at a given node $0$ of the Husimi tree can be telescoped as the sum of the correlations $|G_{0r}|^2$ with all nodes $r$ at fixed distance $r$ from $0$, multiplied by the corresponding values of $\hat{x}_r^{(\nu)}$.
\begin{equation}\label{eq:hatx_path_decomposition}
    \hat{x}_0^{(\nu)} = \sum_{\mathcal{P}: 0 \to r} \left[ \prod_{\ell=1}^r \left | T_{\ell-1}^{(\mu_\ell)}  \right |^2 \right] \hat{x}_r^{(\mu_r)} \, ,
\end{equation}
where the sum runs over the $(kp)^r$ shortest paths $\mathcal{P}$ connecting node $0$ with the nodes at distance $r$ from it, and the product runs over the cliques $\mu_\ell$ traversed by these paths.

Let us therefore consider a reference node of the Husimi tree, labeled as $0$, and assume that $\hat{x}_r^{(\mu_r)} = 1$ on all $(kp)^r$ nodes located at distance $r$ from it. Comparing Eqs.~\eqref{eq:C0r} and~\eqref{eq:hatx_path_decomposition}, the  imaginary part of the linearized Green’s function at the reference node is then
\begin{equation}
\hat{x}_0  \simeq (kp)^r\, |G_{0r}|^2 \, .
\end{equation}
On the other hand, the probability distribution of $\hat{x}_0 $ is also determined, asymptotically at large $r$, by the largest eigenvalue of the linear integral operator defined in Eq.~\eqref{eq:operator_kp}: $\hat{x}_0 \sim \lambda_{1/2}^r$. To leading order, this yields
\begin{equation} \label{eq:C0r1}
C_{0r} \simeq \left( \frac{\lambda_{1/2}}{kp} \right)^{\!r} \, .
\end{equation}
Defining the localization length $\xi_{\rm loc}$ from the exponential decay of the correlation function---after factoring out the trivial geometric contribution $(kp)^{-r}$---we finally obtain
\begin{equation} \label{eq:C0r2}
C_{0r} \propto (kp)^{-r} e^{-r / \xi_{\rm loc}}
\;\; \Longrightarrow \;\;
\xi_{\rm loc} = - \frac{1}{\ln \lambda_{1/2}} \, ,
\end{equation}
as anticipated in the main text and in Eq.~\eqref{eq:loc_len}.

Equivalently, and more formally, the correlation function $C_{0r}$, Eq.~\eqref{eq:C0r}, can be written as $|G_{rr}|^2$ times the action of the same integral operator applied $r$ times, starting from an equilibrium initial condition for $\chi_0^{(\mu_1)}$ governed by the stationary cavity equations. This integral operator is exactly the one defined in Eq.~\eqref{eq:operator_kp}, which governs the evolution of the imaginary part of the Green’s function under iteration in the linearized regime, divided by the geometric factor $(kp)$~\cite{parisi2019anderson}. Note that in Eqs.~\eqref{eq:C0r1} and~\eqref{eq:C0r2} we have omitted a subleading $r^{-3/2}$ correction, originating from the integration over the continuous part of the spectrum of the integral operator~\eqref{eq:operator_kp} (see Refs.~\cite{zirnbauer1986anderson,zirnbauer1986localization} for details).


\section{Numerical solution of the self-consistent recursion cavity equations via Population dynamics}\label{app:num}

As explained in the main text, once averaged over all disorder realizations,  
Eq.~\eqref{eq:cavity} must be interpreted as a self-consistent integral equation for the joint probability distribution of the real and imaginary parts of the cavity Green’s functions:
\begin{widetext}
\begin{equation} \label{eq:PG}
    P(G) = \int \prod_{\mu=1}^k \left [ \prod_{i=1} dG_i^{(\mu)} P(G_i^{(\mu)}) \right ] \delta \! \left( G - \left( E + {\rm i} \eta - \epsilon_0 + t^2 \sum_{\mu \in \partial_0 \setminus \nu}\frac{\sum_{l \in \partial_{\mu} \setminus 0} \frac{G_l^{(\mu)}}{1+tG_l^{(\mu)}}}{1-t\sum_{l \in \partial_\mu \setminus 0} \frac{G_l^{(\mu)}}{1+tG_l^{(\mu)}}} \right)^{\!\!-1} \right) \, .
\end{equation}
\end{widetext}

This equation can be solved very efficiently and with arbitrary numerical precision using the so-called population dynamics algorithm~\cite{mezard2009information,tikhonov2019critical}, described in detail below. 

The method approximates the probability distribution $P(G)$ by the empirical distribution of a large pool of $\Omega$ complex cavity fields ${G_\alpha}$, $P(G) \simeq \Omega^{-1} \sum_{\alpha=1}^{\Omega} \delta(G - G_\alpha)$.  Each iteration consists in updating the entire population: for every element in the pool, a random on-site energy $\epsilon$ is drawn uniformly from $[-W/2,W/2]$, and $kp$ other elements are randomly selected to play the role of the neighboring cavity variables $G_l^{(\mu)}$ appearing in the $\delta$ function of Eq.~\eqref{eq:PG}. A new value of $G$ is computed from this recursion and replaces a randomly chosen element of the pool. This procedure is iterated until the empirical distribution becomes stationary, which can be monitored by tracking the convergence of the moments of $P(G)$.

Once the distribution of cavity Green’s functions has converged, the probability distribution of the full on-site Green’s function in the presence of all neighbors can be obtained in a similar way by implementing Eq.~\eqref{eq:G}. Observables such as the typical local density of states or the inverse participation ratio are then computed as empirical averages over the stationary population.

The accuracy of the method is entirely controlled by the population size $\Omega$~\cite{tikhonov2019critical}. Because $P(G)$ may exhibit broad, power-law tails---especially near criticality---finite-$\Omega$ effects can be substantial, and convergence to the $\Omega \to \infty$ asymptotic limit may be slow (typically, logarithmically slow in $\Omega$). As emphasized in Refs.~\cite{tikhonov2019critical,tonetti2025testing}, particular care is therefore required when extrapolating population dynamics results to large $\Omega$. In the following, we apply the population dynamics method to determine the critical disorder, the mobility edge, and several key observables of the Anderson model on the Husimi tree, using large populations.

\subsection{The localization transition and the mobility edge}

Since in the localized phase the imaginary part of the Green’s function vanishes with the regulator $\eta$, the most direct and transparent way to determine the localization threshold is through a linear stability analysis of Eq.~\eqref{eq:cavity} with respect to a small imaginary component. Introducing the variables $\chi_i^{(\mu)} = G_i^{(\mu)}/(1 + t\, G_i^{(\mu)})$, writing $\chi_i^{(\mu)} = x_i^{(\mu)} + i \eta\, \hat{x}_i^{(\mu)}$, and expanding the recursion up to first order in $\eta$ leads to Eqs.~\eqref{eq:Xreal_self} and \eqref{eq:Ximaginary_self}. These equations define a self-consistent integral equation for the joint probability distribution of $x_i^{(\mu)}$ and $\hat{x}_i^{(\mu)}$.

As discussed in the main text and in Appendix~\ref{app:operator} above, the linear stability of this system is governed by the largest eigenvalue of the linear integral operator \eqref{eq:operator_kp} evaluated at $\beta = 1/2$. This eigenvalue is directly related to the Lyapunov exponent controlling the exponential growth (in the delocalized phase) or exponential decay (in the localized phase) of the typical imaginary part of the Green’s function under iteration:
\begin{equation}
\ln \lambda_{1/2} = \lim_{n_{\rm iter}\to\infty} \frac{\langle \ln \hat{x} \rangle}{n_{\rm iter}} \, .
\end{equation}
To locate the Anderson transition, we solve the linearized equations [Eqs.~\eqref{eq:Xreal_self} and \eqref{eq:Ximaginary_self}] using the population dynamics algorithm and track the asymptotic behavior of the ratio $\langle \ln \hat{x} \rangle / n_{\rm iter}$ as the number of iterations increases. The critical disorder is identified as the point where our estimate of $\ln \lambda_{1/2}$ changes sign and crosses zero. The results obtained with this procedure are shown in Fig.~\ref{fig:W_c}(a) at $E=0$, for varying $k$ and $p$ while keeping the total degree fixed to $z=(k+1)p=12$, for a large population of size $\Omega = 2^{28}\simeq 2.7 \times 10^8$. The critical disorder extracted from the crossing of $\ln \lambda_{1/2}$ with the $x$ axis is then reported in Fig.~\ref{fig:W_c}(b) for $z=12$ and $36$, again at $E=0$.

\begin{figure}
    \centering
    \includegraphics[width=1\linewidth]{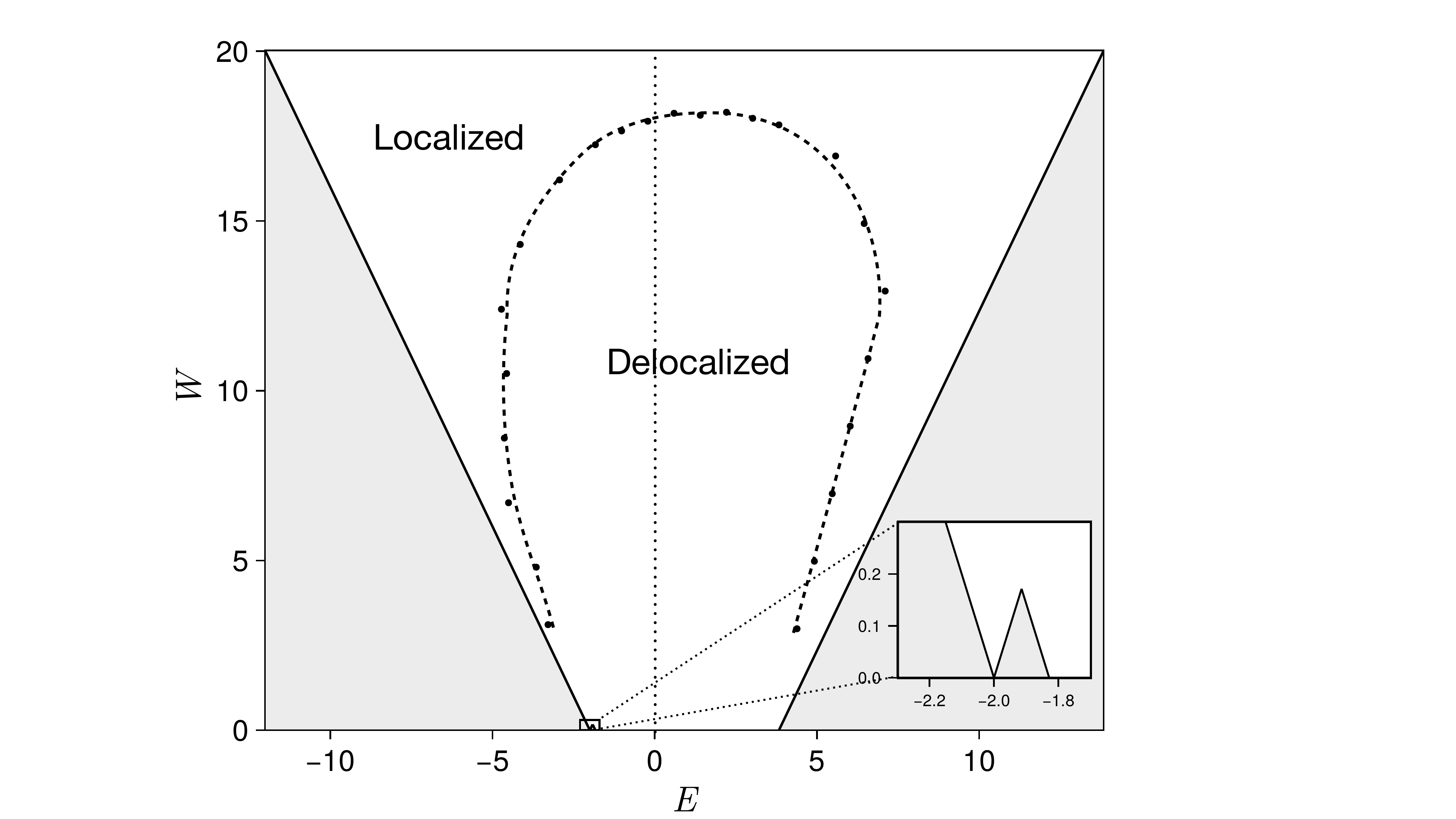}
    \caption{Phase diagram for Anderson localization on Husimi trees for $k = 1$ and $p = 2$. The solid lines mark the theoretical disorder-broadened spectral boundaries~\eqref{eq:edges}. The shaded regions correspond to energies outside the spectral support. Numerical estimates of the mobility edge (circles), obtained from a population of $\Omega = 10^7$, are interpolated by the dashed line. States below this boundary are extended, while those above are localized. The dotted line indicates the $E = 0$ vertical axis, on which all the analysis presented in this work is performed. The inset shows a magnified view of the forbidden region at small disorder, where the band resulting from the broadening of the isolated eigenvalues and the band corresponding to the continuous part of the spectrum have not yet merged.} 
    \label{fig:mobility_edge}
\end{figure}

To determine the position of the mobility edge and obtain the phase diagram, we apply this method across a range of energies $E$ and disorder strengths $W$. The resulting phase diagram for a representative case, $k=1$ and $p=2$, is shown in Fig.~\ref{fig:mobility_edge}. The solid lines mark the disorder-broadened spectral boundaries of the spectrum. These boundaries follow from the edges of the zero-disorder density of states~\eqref{eq:rhoEtot} convolved with the uniform disorder distribution. Both the continuous part of the spectrum, with edges given by Eq.~\eqref{eq:E_lim}, and the $\delta$ peak at energy $-t (k+1)$ are broadened by disorder. The spectral edges are therefore given by
\begin{equation} \label{eq:edges}
\begin{aligned}
\lambda_-(W) & = - t (k+1) -\frac{W}{2} \, , \qquad \\
\lambda_+(W) & = t (p+1) + 2 t \sqrt{kp} + \frac{W}{2} \, .
\end{aligned}
\end{equation}
There is also a forbidden region, shown as a shaded triangle in the zoom in the figure, corresponding to small disorder such that the two bands of the spectrum do not merge, i.e., for energies such that $-t(k+1) + \frac{W}{2} < E < t (p+1) - 2 t \sqrt{kp} - \frac{W}{2}$. Near the spectral edges, the density of states develops an exponentially suppressed Lifshitz tails~\cite{biroli2010anderson}.

Different from the Bethe lattice case, the mobility edge displays an asymmetric shape at small disorder between the left and right parts of the spectrum. This asymmetry reflects the asymmetry of the spectral density at zero disorder, which is strongest for $k=1$, as seen in Fig.~\ref{fig:MP_W0}. However, this asymmetry is progressively reduced upon increasing $W$. The overall shape of the mobility edge is similar to that found on the Bethe lattice~\cite{biroli2010anderson}. Yet, in a parameter regime where $p \gg k$, the gap between the degenerate eigenvalue at $-t(k+1)$ and the continuous band becomes large, and the majority of states belong to the isolated spectral peak. In this case, the low-energy and low-disorder part of the phase diagram could exhibit features different from those shown in Fig.~\ref{fig:mobility_edge}. Since this work focuses on the energy $E=0$ and on very large disorder, where the two spectral bands have completely merged, exploring the low-disorder regime goes beyond the scope of the present work, and is left for future studies.

The mobility edge displayed in Fig.~\ref{fig:MP_W0} has been obtained for a large but fixed population size of $\Omega = 10^7$. A more accurate estimate of the mobility edge (especially at small $W$, within the Lifshitz tails) would require accounting for finite-population effects and extrapolating results obtained at finite $\Omega$ to the $\Omega \to \infty$ limit. For a detailed discussion of this issue, see, for instance, Refs.~\cite{tikhonov2019critical,tonetti2025testing}. In practice, when using populations of size $\Omega = 10^7$, the position of the mobility edge is underestimated by less than $3\%$ (the error could, however, be larger at small disorder, in the Lifshitz-tails region~\cite{tonetti2025testing}). Further improving the accuracy of the mobility-edge location is beyond the scope of the present analysis, whose primary goal is to demonstrate that the overall behavior remains standard---closely resembling that of the infinite, loopless Bethe lattice---despite the structural peculiarities of the Husimi tree.

\subsection{Population dynamics algorithm for the computation of the inverse participation ratio}\label{app:num_IPR}

To compute the inverse participation ratio (IPR) in the localized phase with high numerical accuracy, we adopt the scheme introduced in Ref.~\cite{rizzo2024localized}, which allows one to efficiently extract the IPR from the linearized cavity equations. In fact, the IPR is related to the second moment of $|\mathcal{G}_{ii}|$ via
\begin{equation} \label{eq:IPR}
I_2 = \lim_{\eta \to 0^+} \langle \eta |G_{ii}|^2 \rangle/\langle \mathrm{Im}\,G_{ii} \rangle \, .
\end{equation}
A discussed above, $|G_{ii}|$ is broadly distributed. The power-law tails of its probability distribution would lead, in principle, to divergent expressions for $\langle |G_{ii}|^2 \rangle$. In practice, this divergence is avoided because the power-law behavior is cut off at large values of the imaginary part around $\eta^{-1}$, which marks the limit of validity of the linearized equations. Nevertheless, computing $\langle |G_{ii}|^2 \rangle$ requires accessing the region of large imaginary parts of the Green's function, rather than the region of typical finite values. At first sight, this seems to suggest that the linearized equations might not be useful. Fortunately, this is not the case. To see this, we define
\begin{equation} \label{eq:mii1}
M_{ii} \equiv \frac{1}{G_{ii}} = m_{ii} - {\rm i} \eta \, \hat{m}_{ii} \, , 
\end{equation}
from which  one immediately obtains that 
\begin{equation}
\langle |G_{ii}|^2 \rangle  = \int  Q(m,\hat{m}) \frac{1}{m^2+\hat{m}^2 \eta^2}   \, \de m \, \de \hat{m} \, .
\end{equation}
Similarly, using the fact that ${\rm Im}  G_{ii} = \eta \hat{m}_{ii}/(m_{ii}^2 + \eta^2 \hat{m}_{ii}^2)$, $\langle {\rm Im} G_{ii} \rangle$ is expressed as
\begin{equation}
\langle {\rm Im} G_{ii} \rangle = \int  Q(m,\hat{m}) \frac{\eta {\hat{m}}}{m^2+\hat{m}^2 \alpha^2}   \, \de m\, \de \hat{m} \, ,
\end{equation}
Given that $\hat{m}$ is strictly positive we can make the change of variables $m = \eta \hat{m} x$ that leads to
\begin{eqnarray} 
\langle |G_{ii}|^2 \rangle &=& \int  Q(\eta \hat{m} x  ,\hat{m}) 
\frac{(\eta \, \hat{m})^{-1}}{1 + x^2}   \, \de x \, \de \hat{m} \, , \label{eq:gii} \\
\label{eq:img} \langle {\rm Im} G_{ii} \rangle &=&  \int  Q(\eta \hat{m} x  ,\hat{m}) \frac{1}{1 + x^2}   \, \de x \, \de \hat{m} \, .
\end{eqnarray}
In the $\eta \to 0$ limit we can approximate $Q(\eta \hat{m}  x,\hat{m}) \approx Q(0,\hat{m})$ and perform the integration over $x$ explicitly. Plugging Eqs.~\eqref{eq:gii} and \eqref{eq:img} into Eq.~\eqref{eq:IPR}, one finally obtains~\cite{mirlin1994statistical,rizzo2024localized}:
\begin{equation} \label{eq:IpQ}
    I_2 = \frac{ \int  Q(0  ,\hat{m}) \, \hat{m}^{-1} \,\de \hat{m}}  {\int  Q(0 ,\hat{m})  \, \de \hat{m}} \, .
\end{equation}
To sum up, although the moments of the local Green's functions in the localized phase are controlled by the fact that  $|G_{ii}|$ is $O(1/\eta)$ with probability $O(\eta)$, they can be computed in terms of the {\it typical} values of $M_{ii}$, whose real part is typically $O(1)$ and whose imaginary part is typically $O(\eta)$. The fact that in the localized phase one can use the linearized equations to compute the relevant observables, such as the IPR, facilitates the adoption of highly efficient computational methods that strongly reduce the effect of the finite size of the population.

Crucially, using the equation for the diagonal elements of the Green's functions [Eq.~\eqref{eq:G}] and the definition of $M_{ii}$, Eq.~\eqref{eq:mii1}, the condition $m=0$ in Eq.~\eqref{eq:IpQ} above is equivalent to the condition   
\begin{equation} \label{eq:condition}
    \epsilon_0 = - t^2 \sum_{\mu \in \partial_0}\frac{\sum_{l \in \partial_\mu \setminus 0} \frac{{\rm Re} G_l^{(\mu)}}{1+t {\rm Re} G_l^{(\mu)}}}{1-t\sum_{l \in \partial_\mu \setminus 0} \frac{{\rm Re} G_l^{(\mu)}}{1+t {\rm Re} G_l^{(\mu)}}} \, .
\end{equation}
This condition is satisfied with probability $1/W$ if the absolute value of the right-hand side is smaller than $W/2$, and with probability zero otherwise.

In the population dynamics algorithm, this condition is enforced through a selective sampling procedure. For a given choice of the model parameters $k$, $p$, and $W$, in the localized phase we first run the standard population dynamics algorithm described above to obtain the stationary probability distribution of the cavity Green’s functions in the linearized regime, $P(x,\hat{x})$, corresponding to the solution of Eqs.~\eqref{eq:Xreal_self} and \eqref{eq:Ximaginary_self}. We then extract $(k+1)p$ elements from the population and compute $m_{00}$ and $\hat{m}_{00}$ according to Eq.~\eqref{eq:G}. Next, we evaluate the right-hand side of Eq.~\eqref{eq:condition}, which we denote by $S$ for notational simplicity. If (and only if) $|S|<W/2$, we add $\hat{m}_{00}^{-1}/W$ to the numerator and $1/W$ to the denominator of the expression for $I_2$, in accordance with Eq.~\eqref{eq:IpQ}. This procedure is repeated many times, after which both the numerator and the denominator are normalized by the total number of attempts. The population of cavity Green’s functions is then refreshed by performing a few additional iterations of the standard population dynamics algorithm, and the entire process is iterated until the desired accuracy for $I_2$ is reached.

This approach allows one to compute the IPR with very high precision, even extremely close to the critical point. The IPR data shown in Fig.~\ref{fig:IPR}, as well as the results for the tail exponent $\beta$ in the localized phase presented in Fig.~\ref{fig:critical}(b), were obtained using this method.

\bibliography{Bibliography}

\end{document}